\documentclass[reprint, 
  superscriptaddress,
  amsmath,
  amssymb,
  aps, 
  showkeys,
  ]{revtex4-2}

\usepackage{graphicx}
\usepackage{dcolumn}
\usepackage{bm}

\usepackage[hidelinks]{hyperref}

\begin{document}

\title{Ionisation Calculations using Classical Molecular Dynamics}

\author{Daniel Plummer}
\email{daniel.plummer@physics.ox.ac.uk}
\affiliation{Department of Physics, University of Oxford, UK}

\author{Pontus Svensson}
\affiliation{Department of Physics, University of Oxford, UK}

\author{Dirk O. Gericke}
\affiliation{Centre for Fusion, Space and Astrophysics, Department of Physics, University of Warwick, Coventry CV4 7AL, United Kingdom}
\author{Patrick Hollebon}
\affiliation{AWE, Aldermaston, Reading, Berkshire RG7 4PR, UK}
\author{Sam M. Vinko}
\affiliation{Department of Physics, University of Oxford, UK}
\affiliation{Central Laser Facility STFC Rutherford Appleton Laboratory, Didcot OX11 0QX, UK}

\author{Gianluca  Gregori}
\affiliation{Department of Physics, University of Oxford, UK}

\date{\today}

\begin{abstract}

By performing an ensemble of molecular dynamics simulations, the model-dependent ionisation state is computed for strongly interacting systems self-consistently. This is accomplished through a free energy minimisation framework based on the technique of thermodynamic integration. To illustrate the method, two simple models applicable to partially ionised hydrogen plasma are presented in which pair potentials are employed between ions and neutral particles. Within the models, electrons are either bound in the hydrogen ground state or distributed in a uniform charge-neutralising background.  Particular attention is given to the transition between atomic gas and ionised plasma, where the effect of neutral interactions is explored beyond commonly used models in the chemical picture. Furthermore, pressure ionisation is observed when short range repulsion effects are included between neutrals. The developed technique is general, and we discuss the applicability to a variety of molecular dynamics models for partially ionised warm dense matter.
\end{abstract} 


\maketitle

\section{introduction}

Accurate evaluation of the properties of matter in the warm dense state presents a formidable challenge for both theoretical and computational techniques. Strong ionic coupling hinders analytical calculations, whilst the accurate treatment of quantum-mechanical electrons severely restricts the system size accessible to first-principles calculations. A common heuristic for the first of these two properties is the classical Coulomb coupling parameter which is associated with the ratio of average potential energy to average kinetic energy and, for a fully ionised system, takes the form,
\begin{equation}
  \Gamma = \frac{Z^2 e^2}{4\pi\epsilon_0 a}\frac{1}{k_B T},
\end{equation}
where the charge state of the ions is \(Z\), the system is at temperature \(T\), \(k_B\) is Boltzmann's constant and \(\epsilon_0\) is the dielectric constant. The Wigner-Seitz radius \(\label{eq:wig_seitz}
a=\left({4\pi n}/{3}\right)^{-1/3}
\) is an approximate measure of the interparticle separation and depends on the ionic number density \(n\). Strong coupling implies \(\Gamma \gtrsim 1\). The prominence of quantum statistical effects is quantified with the degeneracy parameter,
\begin{equation}
  \theta = \frac{k_B T}{E_{\rm{F}}},
\end{equation}
which is the ratio of average kinetic energy to Fermi energy \(E_{\rm{F}}\) of the system. Partial degeneracy means the electron kinetic energy is comparable to the Fermi energy, \(\theta \sim 1\). Furthermore partial ionisation, where electrons populate both bound and free states, can occur in Warm Dense Matter (WDM) and is the focus of the current work. For further discussion on the parameters relevant to WDM, see ref. \cite{Bonitz2020}.

A large range of computational and theoretical methods have been applied to WDM \cite{Bonitz2020, Bonitz2024}. Quantum statistical computations such as path integral Monte Carlo can access a host of observables, but are hampered by the so-called fermionic sign problem and are limited to relatively small system size \cite{Bonitz2020, Dornheim2018,Bonitz2024,Filinov2023}. In Molecular Dynamics (MD) approaches, the nuclei are commonly treated as classical point particles, whilst the electron subsystem can be treated to different levels of approximation. Typically, an adiabatic assumption is invoked such that the electron kinetics are decoupled from the ionic motion and in first-principles schemes, the thermal electron distribution is often evaluated in the density functional theory framework \cite{Hutter2009,Bonitz2020,Bonitz2024}. Alternatively, simple One Component Plasma (OCP) models can be employed where electrons are incorporated in effective ion-ion potentials, derived from a model dielectric response \cite{Moldabekov2018b,Moldabekov2019, Blouin2021}. In the original OCP \cite{Brush1966, Hansen1973,Caillol1999,Caillol2010, Onegin2024,Khrapak2014,Baus1980}, the electrons are assumed to form a uniform charge-neutralising background and the ions interact through the bare coulomb potential. The simplicity of the OCP approximations allow simulations of up to millions of particles \cite{Demyanov2022}, and equilibrium properties have been well characterised by many authors, e.g. \cite{Brush1966, Hansen1973,Kahlert2020,Mithen2011}. Semiclassical MD models have also been applied to WDM which include a dynamic representation of the electrons, and therefore simulate the Two Component Plasma (TCP). Prevalent approaches include wave packet molecular dynamics \cite{Klakow1994a, Klakow1994b,Ebeling1997,Knaup1999, Lavrinenko2021,Davis2020, Svensson2023}, quantum statistical potentials \cite{Hansen1978, Filinov2004, Ebeling2006, Calisti2024}, and Bohmian approaches \cite{Larder2019, Campbell2024}. Both OCP and TCP models employ representations of free or weakly bound electrons.

To correctly evaluate the bulk properties of WDM in atomistic simulations, large systems are often required. For example, many particles are needed to characterise transport behaviour in the hydrodynamic limit \cite{Svensson2024,Mithen2011, Schorner2022, Svensson2024b}. Therefore, to extend OCP and TCP methods to partially ionised materials, while retaining the computational efficiency necessary to access substantial system sizes, a chemical picture may be required. In this framework each ionic charge state is considered as a distinct species with appropriate inter-particle potentials, such that the equilibrium charge state distribution must be known to compute both dynamic and static equilibrium properties. Importantly, the distribution will depend on the model being used, so must be computed in a \emph{self-consistent} framework. Ebeling and Militzer attempted such a calculation for a hydrogen plasma wavepacket model by performing simulations to minimise the internal energy as a function of ionisation state at constant entropy, however their approach assumes an ideal entropy term and is therefore limited to weak coupling \cite{Ebeling1997}. In the same work, transitions between bound and free populations were also considered, for which the equilibrium ionisation state becomes an important benchmark. A chemical picture MD approach could also provide a useful reference for TCP modelling strategies that incorporate bound states dynamically such as those discussed in refs. \cite{Calisti2024, Gigosos2018}.

In parallel to the development of MD methods, a rich literature exists surrounding both analytical and semi-empirical models in the chemical picture \cite{Saumon1992,Filinov2023,Juranek2000,Ebeling2017, Winisdoerffer2005, Davletov2023,Zimmerman1980, Ebeling2003, Hummer1988,Potekhin1996}. These models consider distinct chemical species and typically perform variational free energy calculations to obtain their equilibrium value. Various approximations exist, but inter-atomic repulsive interactions are usually treated on the level of hard spheres, which exclude the volume accessible to other species'. The chemical picture is also applied to predict x-ray scattering spectra of dense plasmas through the Chihara decomposition \cite{Chihara1987,Bonitz2024}. Furthermore, the ionisation state is generally a key property in the hydrodynamic and spectroscopic modelling of plasmas \cite{Stanton2016, Chung2005, Callow2023, Ciricosta2012}.

Given these motivations, the current manuscript is devoted to computing exact \emph{model-dependent} ionisation states directly from MD, an approach we do not believe has previously been employed. Specifically, we consider a simple system which is applicable to partially ionised hydrogen and perform a set of excess free energy computations to determine the self-consistent equilibrium ionisation state in the presence of strong coupling. Furthermore, the generalisation to materials with higher atomic numbers is straightforward, provided that realistic inter-particle potentials are supplied. To calculate free energies we devise a framework based on the Thermodynamic Integration (TI) technique \cite{Frenkel2002}. The manuscript consists of three core sections: In section \ref{sec:model}, two models applicable to partially ionised hydrogen plasma are introduced. Following this, in section \ref{sec:free_min_framework}, we develop a free energy minimisation framework to compute the ionisation state of a given model. Section \ref{sec:simulation_results} details the application of
this framework and presents results for the computed ionisation states over a range of densities. Finally, conclusions are drawn in section \ref{sec:conclusion}.
\section{Hydrogen models} \label{sec:model}

We consider a partially ionised and charge-neutral coulomb system in a box of volume \(V = L^3\). The box is subject to periodic boundary conditions and contains \(N\) electrons, and an equal number of protons. A chemical picture is employed, where free electrons form a uniform background charge, and each bound electron is spatially localised on a specific ion. The system may therefore be partitioned into \(N_{\rm{i}}\) ions and \(N_{\rm{n}}\) neutrals, such that \(N = N_i + N_n\). The general expression for the Hamiltonian used in the MD simulations is 
\begin{equation} \label{eq:MD_hamiltonian}
  \mathcal{H}^{(\rm{MD})} = \sum_k^{N_{\rm{i}}} \frac{{(\mathbf{p}^{\rm{i}}_k)^2}}{2m_{\rm{i}}} + \sum_l^{N_{\rm{n}}} \frac{{(\mathbf{p}^{\rm{n}}_l)^2}}{2m_{\rm{n}}} + \mathcal{U}^{(\rm{MD})},
\end{equation}
where \(m_{\rm{i}}\) is the ion mass, and the neutral mass \(m_{\rm{n}} = m_{\rm{i}} + m_{\rm{e}}\) is the sum of electron and ion masses. The momentum of the \(k^{\rm{th}}\) ion is \(\mathbf{p}^{\rm{i}}_k\) and the \(l^{\rm{th}}\) neutral is \(\mathbf{p}^{\rm{n}}_l\). The potential energy function \(\mathcal{U}^{(\rm{MD})}\) determines the interactions of the system and depends on the configuration of the particles -- modifications to this function serve as the central object for the Thermodynamic Integration computation given in section \ref{sec:TI}. In the following subsections, the potentials used to model the particle interactions are introduced.

\subsection{Bound State Coulomb Model} \label{sec:neutral_model}
In the first model considered, which is referred to as the Bound State Coulomb (BSC) model, each bound electron is modelled as a frozen single-particle wavefunction given by (atomic units are used unless otherwise specified):
\begin{equation}\label{eq:1s}
  \psi_{\rm{1s}} (\mathbf{x}) = \frac{1}{\sqrt{\pi}}\exp\left(-|\mathbf{x}|\right),
\end{equation}
which is equivalent to the ground state of an isolated hydrogen atom, up to a global phase. A neutral particle is configured by superposing the electron onto an ion at position \(\mathbf{r}^{\rm{n}}_l\) giving access to the charge density of a single neutral,
\begin{equation} \label{eq:neutral_density}
  \rho^{\text{n}}(\mathbf{x}, \mathbf{r}^{\rm{n}}_l) = \delta^3(\mathbf{x} - \mathbf{r}^{\rm{n}}_l) - |\psi_{\text{1s}}(\mathbf{x} - \mathbf{r}^{\rm{n}}_l)|^2.
\end{equation}
The subscript index \(l\) labels a given neutral particle. Alternatively, in the ionised case, the ion at position \(\mathbf{r}_k^{\rm{i}}\) and corresponding free electron can be combined to give the charge density of a pseudo-ion: 
\begin{equation}\label{eq:ion_density}
  \rho^{\rm{i}}(\mathbf{x}, \mathbf{r}_k^{\rm{i}}) = \delta^3(\mathbf{x} - \mathbf{r}_k^{\rm{i}}) - 1/{V}.
\end{equation}
Similarly, the subscript index \(k\) labels a pseudo-ion particle. The interactions are written in terms of the pseudo-ion, such that the electron background acts to modify the heavy-particle interactions. In this section, we henceforth consider the system to be composed of \(N_{\rm{n}}\) neutrals and \(N_{\rm{i}}\) pseudo-ions  with charge distributions given by equations \eqref{eq:neutral_density} and \eqref{eq:ion_density} respectively.  To compute the potential of the system, Poisson's equation is solved for these charge distributions, subject to periodic boundary conditions. Specifically, the ion-neutral (in) and neutral-neutral (nn) interaction potentials are computed with,
\begin{equation} \label{eq:integral}
  U^{\rm{in}/\rm{nn}}(|\mathbf{r}^{\rm{i/n}}_k-\mathbf{r}^{\rm{n}}_l|) = \int_{V}\int_{V} \frac{\rho^{\rm{i}/\rm{n}}(\mathbf{x},\mathbf{r}^{\rm{i/n}}_k)\rho^{\rm{n}}(\mathbf{x}', \mathbf{r}^{\rm{n}}_l)}{|\mathbf{x} - \mathbf{x}'|}d\mathbf{x}'d\mathbf{x},
\end{equation}
which acts between distinct particles. The model admits short range pair potentials which are a function of the radial distance \(r\) between particles and determine the neutral dynamics,
\begin{align} \label{eq:ion_neutral_pot}
  U^{\text{in}}_{\rm{pair}}(r) &= \frac{\text{e}^{-2r}}{r}\left(1 + r\right), \\
  U^{\text{nn}}_{\rm{pair}}(r) &= \frac{\text{e}^{-2r}}{r}\left(1 + \frac{5}{8}r - \frac{3}{4}r^2 - \frac{1}{6}r^3\right).
\end{align}
These potentials resemble those used in ref. \cite{Ebeling1997}, and the neutral interactions are screened on the length scale of the bound states, as expected. There is also a volume-dependent contribution from the neutrals interacting with the free electron background charge, 
\begin{equation} \label{eq:ion_neutral_constant}
  U_0^{\text{in}} =  -\frac{2\pi N_{\rm{i}}N_{\rm{n}}}{V},
\end{equation}
which directly arises from computing \(U^{\rm{in}}\) in equation \eqref{eq:integral}. To ease the presentation of the TI method in section \ref{sec:TI} the total interaction energy between different species are explicitly defined. The total energy between pseudo-ions and neutrals is
\begin{equation} \label{eq:total_ion_neutral_energy}
  V_{\rm{in}} = U_0^{\rm{in}} + \sum_{k,l}\sum_{\mathbf{n}} U^{\rm{in}}_{\rm{pair}}(|\mathbf{r}^{\rm{i}}_k - \mathbf{r}^{\rm{n}}_l + \mathbf{n}L|),
\end{equation}
where the first sum over \(k,l\) is over all ion-neutral pairs and the second sum is over the vector of integers \(\mathbf{n} = (n_x, n_y, n_z) \in \mathbb{Z}^3\). The total neutral-neutral energy is similarly given by the following double summation over neutral positions,
\begin{equation} \label{eq:total_neutral_neutral_energy}
  V_{\rm{nn}} = \frac{1}{2}\sum_{l,m} \sum_{\mathbf{n}}' U^{\rm{nn}}_{\rm{pair}}(|\mathbf{r}^{\rm{n}}_l - \mathbf{r}^{\rm{n}}_m + \mathbf{n}L|).
\end{equation}
The prime notation indicates that \(\mathbf{n} = (0,0,0)\) term is omitted for \(l=m\) i.e. there is no self interaction in the central box, but particles do interact with their own periodic images. The factor of one half in equation \eqref{eq:total_neutral_neutral_energy} corrects for double counting.
\subsection{OCP Interactions}

The remaining component are the pseudo-ion interactions which form a One Component Plasma (OCP) subsystem and  can be similarly decomposed as pair and constant terms \cite{Brush1966, Onegin2024, Fraser1996}. The pair potential
\begin{equation} \label{eq:ewald_pair}
  U^{\rm{ii}}_{\rm{pair}}(\mathbf{r}) = v_1(\mathbf{r}) + v_2(\mathbf{r})
\end{equation}
is computed under the Ewald method and therefore split into two convergent summations \cite{Rapaport2004}. The first term represents real space contributions to the energy and is given by
\begin{equation} \label{eq:ewald_short}
  v_1(\mathbf{r}) = \sum_{\mathbf{n}}\frac{\rm{erfc}(g |\mathbf{r} + \mathbf{n}L|)}{|\mathbf{r} + \mathbf{n} L|},
\end{equation}
where \(g\) is the Ewald parameter, chosen to optimise numerical performance \cite{Peterson1995}, and \(\rm{erfc}(\cdot)\) is the complementary error function. The second term in equation \eqref{eq:ewald_pair} represents reciprocal space contributions to the pairwise energy,
\begin{equation}\label{eq:ewald_long}
  v_2(\mathbf{r}) = \frac{1}{V}\sum_{\mathbf{k}\neq0}\frac{4\pi}{k^2}{e^{{-k^2/4g^2}}}\cos(\mathbf{k}\cdot\mathbf{r}),
\end{equation}
and runs over non-zero k-vectors where \(\mathbf{k} = 2\pi \mathbf{n}/L\). Furthermore there is a constant energy contribution comprised of 4 terms \cite{Onegin2024},
\begin{equation} \label{eq:ewald_const}
\begin{split}
  U_{0}^{\rm{ii}} = \frac{N_i}{2}\sum_{\mathbf{n} \neq 0} \frac{{\rm{erfc}}(gnL)}{\mathbf{n}L} + \frac{N_i}{2}\sum_{\mathbf{k} \neq 0} \frac{4\pi}{k^2}e^{-k^2/4g^2} \\ -N_{\rm{i}}\frac{g}{\sqrt{\pi}} - N_{\rm{i}}^2\frac{\pi}{2 g^2 V}.
\end{split}
\end{equation}
The OCP interactions fully account for the ion-background and background-background contributions of the model \cite{Brush1966}. Generally, the presence of the uniform free electron background provides a regularisation to the ion-ion interactions, which would otherwise formally diverge, and contributes a finite energy offset to all inter-species potentials involving pseudo-ions, e.g. equation \eqref{eq:ion_neutral_constant}. These constant energy terms do not affect the dynamics but do depend on the number of particles, and therefore are particularly relevant for comparing free energies at different ionisation states. Similarly, electron background energy contributions have been discussed in the context of phase diagram calculations \cite{Blouin2021}. The total OCP energy consists of the constant term and a sum over pairs of pseudo-ions,
\begin{equation}
  V_{\rm{ii}} = U^{\rm{ii}}_0 + \sum_{j < k}  U^{\rm{ii}}_{\rm{pair}}(|\mathbf{r}^{\rm{i}}_j - \mathbf{r}^{\rm{i}}_k|),
\end{equation}
The OCP pair potential has the periodic boundary conditions included and therefore no sum over periodic images is required, in contrast to equations \eqref{eq:total_ion_neutral_energy} and \eqref{eq:total_neutral_neutral_energy}.

\subsection{Short Range Repulsion model} \label{sec:SRR_potential}

In plasma models that apply a chemical picture, hard sphere potentials are often used, either as a reference potential with known properties e.g. \cite{Saumon1992}, or directly e.g. \cite{Ebeling2003}. The justification of this second class of potentials is left to section \ref{sec:SRR_TI}. To investigate an analogous effect, we introduce a second model involving a Short Range Repulsion (SRR) between neutral particles, noting that similar potentials have been previously introduced to model bound electron repulsion effects for partially ionised plasmas \cite{Wunsch2009,Vorberger2013}. A third potential,
\begin{equation} \label{eq:SRR_potential}
  U_{\rm{SRR}}^{\rm{nn}} = \epsilon\left(\frac{1}{r}\right)^{12},
\end{equation}
is applied in combination with the BSC model, that acts solely between neutral pairs. In equation \eqref{eq:SRR_potential}, \(\epsilon\) is the overall constant determining the strength of the interaction and is set by equating {\(U_{\rm{\rm{SRR}}}^{\rm{nn}}(2r_*) = (3/2)k_B T\), to give an effective radius \(r_*\).} The total energy due to SRR interactions is
\begin{equation}
  V_{{\rm{SRR}}} = \sum_{l < m}\sum_{\mathbf{n}}'{U^{\rm{nn}}_{\rm{SRR}}}(|\mathbf{r}^{\rm{n}}_l - \mathbf{r}^{\rm{n}}_m + \mathbf{n}L|),
\end{equation}
where the sum runs over all pairs of neutral particles. The potentials discussed thus far have been implemented as a new pairstyle into the \textsc{LAMMPS} codebase \cite{LAMMPS}, in which all subsequent simulations are performed. The MD potential is given by
\begin{equation}
  \mathcal{U}^{\rm{(MD)}} = V_{\rm{ii}} + V_{\rm{in}} + V_{\rm{nn}} + V_{\rm{SRR}},
\end{equation}
where for the BSC model, the SRR term is discarded, or equivalently, \(\epsilon = 0\). Having defined the two hydrogen models, the ionisation calculation methodology will now be introduced.

\section{Free Energy Minimisation Framework} \label{sec:free_min_framework}

In any model which distinguishes between bound and free electrons, the ionisation state is key quantity of interest. In hydrogenic systems, the ionisation state takes a particularly simple form as the ratio of the number of ions to the total number of electrons \(\bar{z} = N_{\rm{i}}/N\). A system in thermodynamic equilibrium and held at constant density \(n=N/V\) and temperature \(T\) will exhibit a minimum in the Helmholtz free energy:
\begin{equation}\label{eq:minimisation}
  \left(\frac{\partial f}{\partial \bar{z}}\right)_{n,T} = 0.
\end{equation}
Equation \eqref{eq:minimisation} is valid under the constraint of charge neutrality and \(f = F/Nk_B T\) is the reduced free energy per particle. The equilibrium ionisation state may therefore be found by computing the free energy across a set of artificial ionisation states and performing a minimisation procedure. These considerations apply to any chemical picture of hydrogen, and are readily generalisable to heavier elements \cite{Ebeling2017,Winisdoerffer2005,Kumar2021}, where different charge states will be present. To our knowledge, this approach has not been employed in MD applications where, given a set of interparticle potentials, the effect of strong coupling can be treated exactly. The methodology to compute the free energy is detailed in the following section.

\subsection{Thermodynamic Integration} \label{sec:TI}

\begin{figure*}
  \centering
  \includegraphics[width=\linewidth]{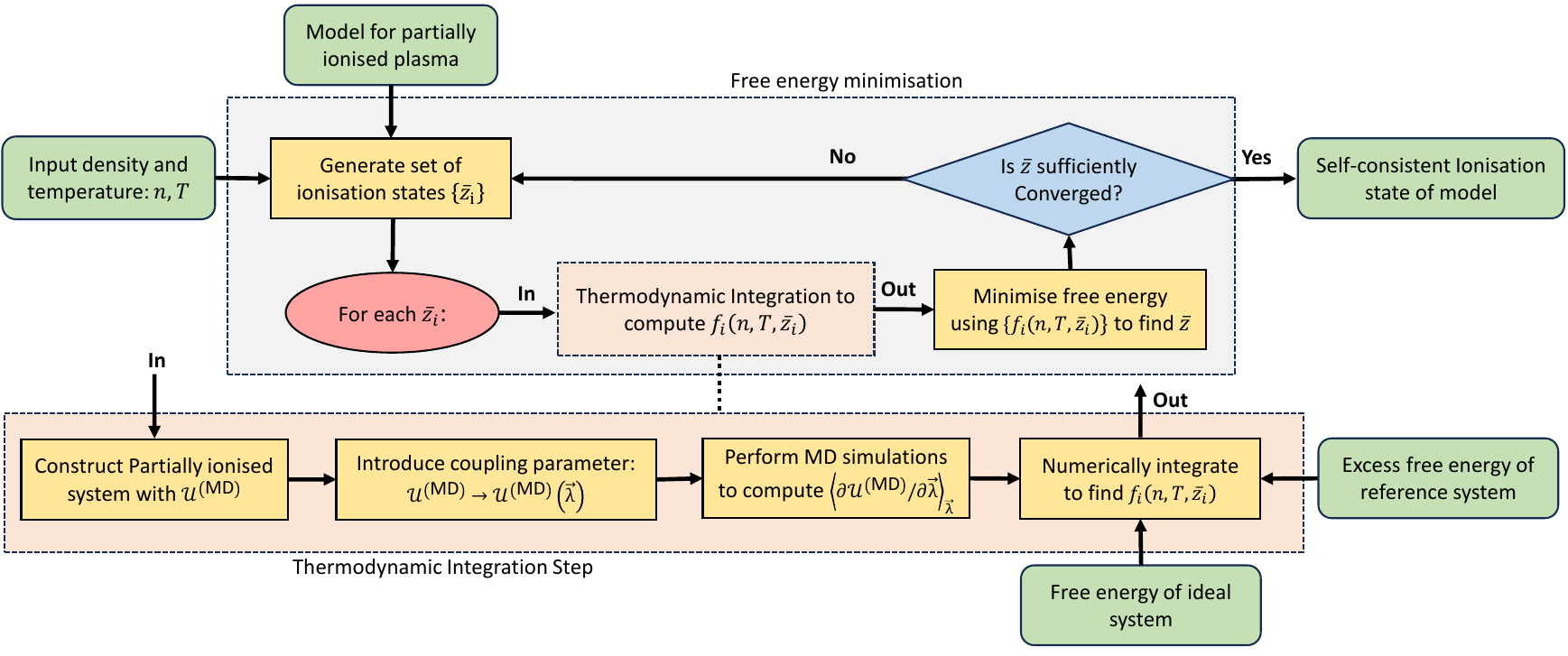}
  \caption{\label{fig:free_min_framework} Schematic depicting the free energy minimisation framework, including the Thermodynamic Integration step (enlarged pink box). The framework requires external input to the calculation (green boxes) and a set of computation steps (yellow boxes), while the red oval represents an iteration over ionisation states and the blue diamond a decision. The output is the self-consistent ionisation state of the input model which is also given in a green box. The schematic applies to both models presented in the text, where for the SRR model, the reference system is chosen to be the BSC model.}
  \label{fig:enter-label}
\end{figure*}

The Helmholtz free energy depends on the accessible volume of the underlying phase space, and is not directly accessible in MD simulations. For this reason, we have performed a multistep Thermodynamic Integration (TI) to compute the free energy across different ionisation states \cite{Frenkel2002}. A coupling vector \(\vec{\lambda}\) may be introduced into the MD potential,
\begin{equation} \label{eq:potential_with_coupling}
   \mathcal{U}^{(\rm{MD})} \rightarrow \mathcal{U}^{(\rm{MD})}(\vec{\lambda})
\end{equation}
in a currently unspecified manner. Two important cases are imposed: When the coupling vector is the null vector, \(\vec{\lambda} = \vec{0}\), the system is noninteracting (ideal) and when the coupling vector is the unit vector, \(\vec{\lambda} = \vec{1}\), the interactions match the those of the model i.e. the interactions of the \emph{target} system. The introduction of coupling parameters allows the excess free energy to be written as an integral over ensemble averages evaluated over different coupling strengths \cite{Frenkel2002}:
\begin{equation} \label{eq:TI_integration_2d}
  f^{\rm{ex}}(n, T, \bar{z}) = \frac{1}{Nk_B T}\int_{\vec{0}}^{\vec{1}}  d\vec{\lambda} \cdot \left\langle\frac{\partial \mathcal{U}^{(\rm{MD})}(\vec{\lambda})}{\partial{\vec{\lambda}}} \right\rangle_{\vec{\lambda}}.
\end{equation}
The notation \(\langle \cdot \rangle_{\vec{\lambda}}\) denotes an ensemble average computed for a system at the given coupling vector. In the context of MD, this consists of evaluating time averages over trajectories generated by the Hamiltonian of the system with the potential interactions modified by the coupling vector. Given these trajectories, the components of the vector defined by \(\partial \mathcal{H} / \partial \vec{\lambda}\) may be evaluated over them. Therefore a set of simulations are required to numerically evaluate the integral for each ionisation state. A schematic for the full minimisation framework is presented in figure \ref{fig:free_min_framework}, including the TI step. Given that equation \eqref{eq:TI_integration_2d} computes the excess free energy for a set of particle interactions the ideal free energy of the model is required as input for the calculation.

\subsection{Ideal Free Energy}

In line with the BSC model, each hydrogen atom is in its spin-degenerate ground state with binding energy \(I \approx 13.59\) \(\rm{eV}\). Free electrons have two spin states and follow Fermionic statistics. Given that the ideal system is noninteracting, the free energy of the individual species combine linearly,
\begin{equation} \label{eq:ideal_free}
  f^{\rm{id}}(n, T, \bar{z}) = f^{\rm{id}}_i + f^{\rm{id}}_e + f^{\rm{id}}_n,
\end{equation}
where the free energy of the ions \(f_i^{\rm{id}}\) and neutrals \(f_n^{\rm{id}}\) are respectively, 
\begin{align}
  &f_i^{\rm{id}} =  \bar{z}\log[\bar{z}n\Lambda_i^3] - \bar{z}, \\
  &f_n^{\rm{id}} = \bar{x}\log\left[\frac{1}{2}\bar{x} n\Lambda_n^3\right] - \bar{x}\frac{I}{k_B T} - \bar{x},
\end{align}
where \(\Lambda_{\alpha}\) is the thermal de-Broglie wavelength of the species \(\alpha\) and \(\bar{x}=1-\bar{z}\) is the neutral fraction. The electrons follow a Fermi-Dirac distribution over free particle states, and their free ideal free energy is given by

\begin{equation}\label{eq:free_fermi_electron}
  f^{\rm{id}}_{e} = -\frac{2}{3}\frac{\mathcal{F}_{3/2}\left(\eta_e\right)}{\mathcal{F}_{1/2}\left(\eta_e\right)} + \eta_e,
\end{equation}
where \(\eta_e = \mu_e/k_B T\) is the reduced chemical potential and \(\mathcal{F}_j\) is a Fermi-Dirac integral of order \(j\). The chemical potential is computed through the normalisation condition \(\mathcal{F}_{1/2}\left(\eta_e\right) = {\bar{z}n \Lambda_e^3}/{2}\).
On application of the free energy minimisation condition, equation \eqref{eq:minimisation}, to  the ideal free energy, equation \eqref{eq:ideal_free}, the ideal Saha equation can be recovered in the limit where the free electrons are non-degenerate, \(\theta \ll 1\). Further details are given in appendix \ref{app:saha_equation}. There are different avenues by which excited atomic states could be included within the calculation, either by assuming a parameterised distribution of states in the atom, and modifying the interaction accordingly, or introducing each excited state as another independent species. These extensions pose no fundamental problems to the calculation, but increase the complexity, and are beyond the scope of the current manuscript.

\subsection{Bound State Coulomb Model Free Energy Calculation}

To compute the free energy of the BSC model, 2 coupling parameters \(\vec{\lambda} = (\lambda_1, \lambda_2)\) are introduced into the MD potential. The first entry \(\lambda_1\) controls the strength of pseudo-ion interactions, while \(\lambda_2\) modifies all neutral interactions such that
\begin{equation}
  \mathcal{U}^{\rm{(MD)}}(\vec{\lambda}) = \lambda_1 V_{\rm{ii}} + \lambda_2(V_{\rm{in}} + V_{\rm{nn}}).
\end{equation}
The intermediate state \(\vec{\lambda} = (1,0)\) constitutes an ideal gas of neutrals immersed in an interacting OCP. Given that extensive equation of data exists for the OCP it serves as the \emph{reference} system for the calculation, with free energy \(f^{\rm{OCP}}\) which solely depends on the effective coulomb coupling parameter which itself is a function of the ionisation state, density and temperature. We provide our own parameterisation of data from Caillol \cite{Caillol1999,Caillol2010}, specified in appendix \ref{app:OCP_data}. The OCP data involves integrating over a weakly-coupled system with long-range potentials, where the Debye length is necessarily larger than the box size. In this case, the finite size error is sizeable and many more particles have to be used to estimate the interaction energy in the thermodynamic limit \cite{Caillol2010}, further discussion is given in appendix \ref{app:OCP_data}. By using the OCP as the reference system in the calculation, we overcome these issues in a computationally efficient manner. To compute the free energy of the target system, \(f^{\rm{BSC}}(n, T, \bar{z})\), we evaluate,
\begin{equation} \label{eq:TI_neutral}
  f^{\rm{BSC}} = f^{\rm{OCP}} + \frac{1}{Nk_BT}\int_0^1 d\lambda_2 \langle V_{\rm{in}} + V_{\rm{nn}} \rangle_{(1, \lambda_2)},
\end{equation}
which follows from equation \eqref{eq:TI_integration_2d}, over different ionisations states.

\subsection{Short Range Repulsion Free Energy Calculation}
To compute the free energy of the SRR model, a third coupling parameter is introduced that controls the strength of the additional repulsive interactions,
\begin{equation}
  \mathcal{U}^{\rm{(MD)}}(\vec{\lambda}) = \lambda_1 V_{\rm{ii}} + \lambda_2(V_{\rm{in}} + V_{\rm{nn}}) + (\lambda_3)^{12} V_{\rm{SRR}}.
\end{equation}
Introducing this parameter to the power of \(12\), mirroring the dependence of the SRR potential, equation \eqref{eq:SRR_potential}, on radial separation, avoids singularities when evaluating the ensemble average of \(\partial V_{\rm{SRR}}/\partial \lambda_3\) \cite{Beutler1994}. The free energy is calculated by a natural extension of formula \eqref{eq:TI_integration_2d}, where the BSC model now acts as the reference system:
\begin{equation} \label{eq:TI_SRR}
  f^{\rm{SRR}} = f^{\rm{BSC}} + \frac{1}{Nk_BT}\int_0^1 d\lambda_3 \langle 12\lambda_3^{11} V_{\rm{SRR}} \rangle_{(1, 1, \lambda_3)},
\end{equation}
which is an application of equation \eqref{eq:TI_integration_2d} where the coupling vector now has 3 entries \(\vec{\lambda} = (\lambda_1, \lambda_2, \lambda_3)\). Computational expense is reduced by choosing the reference system as the BSC model, in which there are already soft repulsive interactions present. The TI method benefits from having similar reference and target systems, such that fewer coupling parameter points are needed to accurately resolve the TI curve and estimate the integral in equation \eqref{eq:TI_integration_2d} \cite{Ruiter2021}.

\section{Simulation Results}\label{sec:simulation_results}

In this section simulation results pertaining to the computation of the the ensemble averages in equations \eqref{eq:TI_neutral} and \eqref{eq:TI_SRR} are presented. The subsequent free energy minimisation and ionisation calculation are also given and discussed. All simulations were performed for a constant temperature \(T = 62500 \: \rm{K} \approx 5.38 \: \rm{eV} \) across a range of densities \(r_s = a/a_B = \{4, 3, 2, 1.5, 1, 0.75, 0.5\}\) where \(a_B\) denotes the Bohr radius. These conditions were selected because at this temperature, molecule formation between hydrogen atoms is unlikely while partial ionisation is predicted to occur \cite{Filinov2004, Filinov2023}. Ionisation states were chosen in an iterative scheme to sample the minima in free energy. In each simulation \(N=1024\) heavy particles were initialised from a configuration that minimised the potential energy, thermalised for 2.5 ps to the target temperature by applying a velocity rescaling every 0.5 fs, before trajectory data was collected over 5 ps. For both models, statistics were then gathered over 6 independent runs, and used to produce the thermodynamic integration curves discussed below. This procedure ensures that different local configurations of the canonical ensemble are sampled. Finite size tests were carried out and confirmed sufficient convergence with respect to system size with a relevant example given in appendix \ref{app:finite_size_tests}. At high densities, \(r_s \leq 0.75\), convergence was not reached for the BSC model at low ionisation states. This corresponds to a breakdown of the model and further discussion is provided in section \ref{sec:ionisation_across_density}.

\subsection{Free Energy Minimisation of Bound State Coulomb Model} \label{sec:free_min_hydrogen_model}

\begin{figure}
  \centering
  \includegraphics[width=\linewidth]{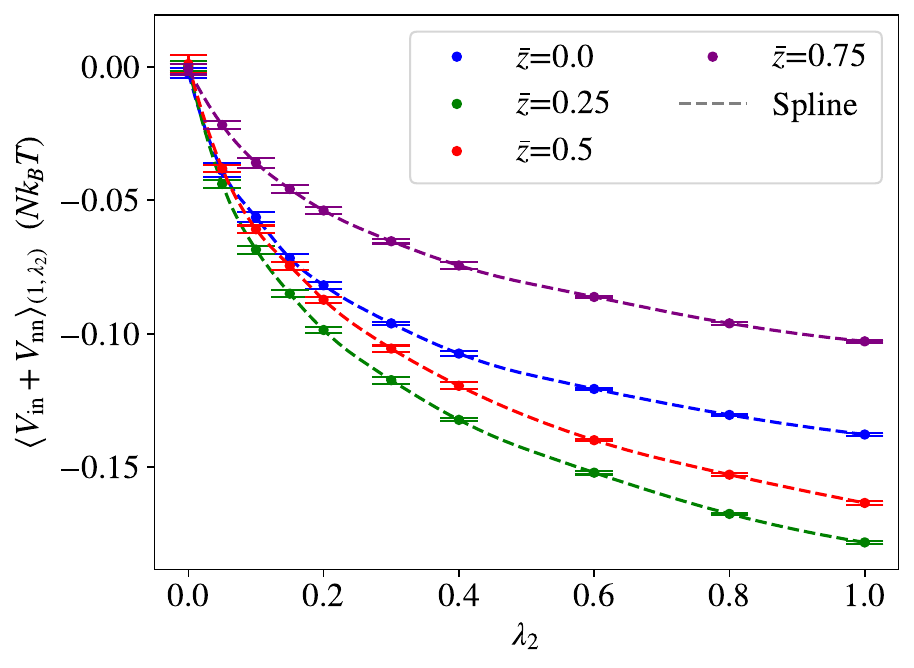}
  \caption{\label{fig:TI_example} Thermodynamic Integration curves for the BSC model at \(r_s = 2\), the integrand in equation \eqref{eq:TI_neutral} is plotted for different values of \(\lambda_2\). For clarity, a subset of ionisation states are displayed. The data was averaged over 6 independent runs, while error bars are the corresponding standard deviations.}
\end{figure}

Example Thermodynamic Integration curves are given in figure \ref{fig:TI_example}, where results are shown for \(r_s = 2\) and selected ionisation states. At \(\lambda_2=0\), all curves start at zero. The neutral trajectories are uncorrelated and therefore sample an integral over the ion-neutral and neutral-neutral potentials, both of which integrate to zero over the box. In the literature this condition is often used to set the zero of energy in periodic systems \cite{Figueirido1995, Caillol1999a}, a convention followed here. As \(\lambda_2\) increases, the curves exhibit an initial sharp decrease, before entering a linear regime. Furthermore, for small \(\lambda_2\), the fluctuations in the ensemble averages are larger. Both effects are caused by trajectories passing close by one another due to the weak interactions. This results in large fluctuations in the magnitude of the potentials when evaluated over these paths, but with sufficient averaging this effect is reduced, as demonstrated in figure \ref{fig:TI_example}, where the corresponding errors are small. A spline is fit to each curve and integrated to find the free energy with equation \eqref{eq:TI_neutral}. Given that the spline fitting is less accurate in regions of larger curvature, the integrand is sampled at finer resolution for lower coupling parameters to provide a more accurate calculation of the free energy. A discussion of the error associated to the integrand sampling is also given in appendix \ref{app:finite_size_tests}.

\begin{figure}
  \centering
  \includegraphics[width=\linewidth]{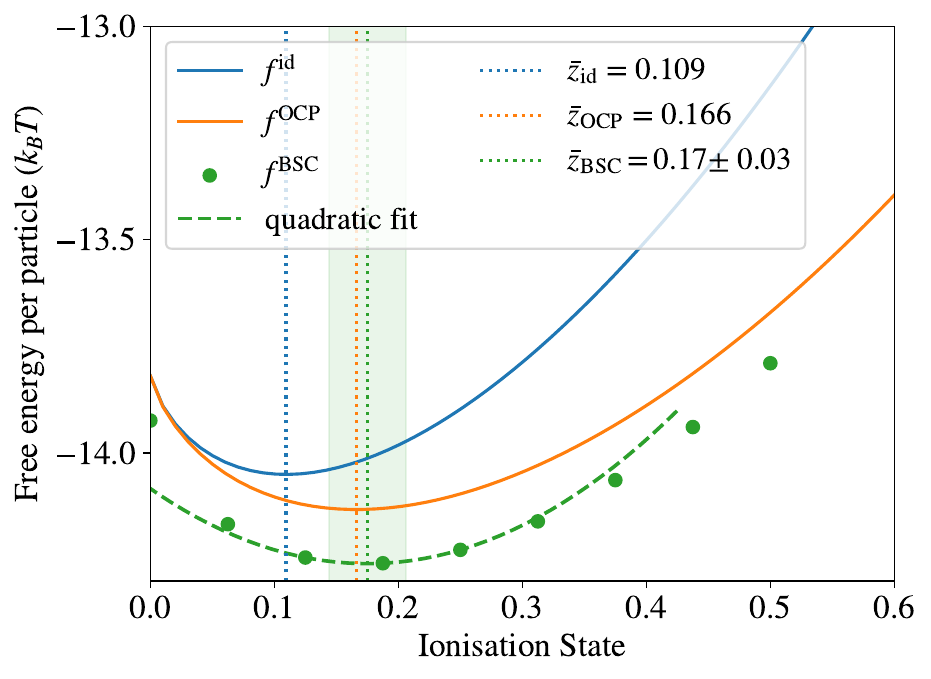}
  \caption{\label{fig:neutral_free_energy_min} BSC Model free energy minimisation for \(r_s=2\). The curves with solid lines correspond to free energies of the ideal (blue, \(\vec{\lambda} = \vec{0}\)), reference (orange, \(\vec{\lambda} = (1,0)\)), and target (purple, \(\vec{\lambda} = \vec{1}\)) system, while the dotted lines show their minima, the uncertainty in the quadratic minimisation is also displayed in the green shaded rectangle. The corresponding values of these minima are given in the legend.}
\end{figure}

Figure \ref{fig:neutral_free_energy_min} shows the corresponding free energy curves for the BSC model, where all sampled ionisation states are included. The ideal case and reference systems are also plotted, which are parameterised functions and are readily minimised. A finite set of data are available to minimise the neutral free energy, and therefore a quadratic curve is fit to the lowest 3 data points to extract the minima. The error mainly depends upon the resolution in ionisation state, which we estimate to be half the interval over which the quadratic curve is fitted. When including the OCP interactions, the collective binding energy between the ions and free electron background \emph{lowers} the free energy at higher ionisation states. Therefore, the ionisation state that minimises the free energy increases. Only ion screening is present and the inclusion of an electronic response would cause further increase, as there are additional negative free energy contributions due to screening \cite{Caillol2000}. The attractive coulomb interactions increasing the ionisation state is a form of ionisation potential depression, and corresponds to a lowering of the energy difference between the bound state and the continuum, a connection which is asserted in appendix \ref{app:saha_equation}. At \(\bar{z}=0\), the OCP and ideal systems have the same free energy, given that no ions are present.

The BSC interactions further lower the magnitude of free energy, and the net effect produces a small positive shift in the ionisation state. The decrease in energy is caused by binding between neutrals and the background and the negative asymptote in the neutral-neutral interaction. The latter is the reason for the decrease in free energy at \(\bar{z}=0\) while at \(\bar{z}=1\) the free energies of the target and reference system necessarily agree. The repulsive effects present in the potentials at small interparticle distances contribute to the small increase in ionisation. 

\subsection{Including Short Range Repulsion Effects} \label{sec:SRR_TI}

Hard sphere potentials have been used many times within the chemical picture literature as computational devices \cite{Filinov2023,Saumon1992,Juranek2000,Zimmerman1980,Ebeling2017,Potekhin1996,Hummer1988, Ebeling2003,Winisdoerffer2005,Zimmerman1980}. The charge-neutral interaction in this approach simply restricts the volume accesible to charged species' which is an analytically tractable mechanism and has therefore found prevalent use. For example, Saumon and Chabrier found that, at high densities, the free energy of their model was lower when fully neutral compared to when fully ionised \cite{Saumon1992}. Given that the system was expected to be fully pressure ionised, an additional hard sphere potential was introduced between neutral species with radius \(2 \rm{a}_0\). Although the model presented here does not include the same framework of internal atomic states, or interatomic interactions, we still observe the same phenomenon. Along similar lines, Potekhin introduced an additional unbinding free energy to account for the inner-core screening effect of overlapping atomic orbitals \cite{Potekhin1996}. Furthermore, in many models in the chemical picture, neutral species' interaction effects are modelled directly as hard spheres \cite{Filinov2023, Zimmerman1980, Ebeling2003}. 

\begin{figure}
  \centering
  \includegraphics[width=\linewidth]{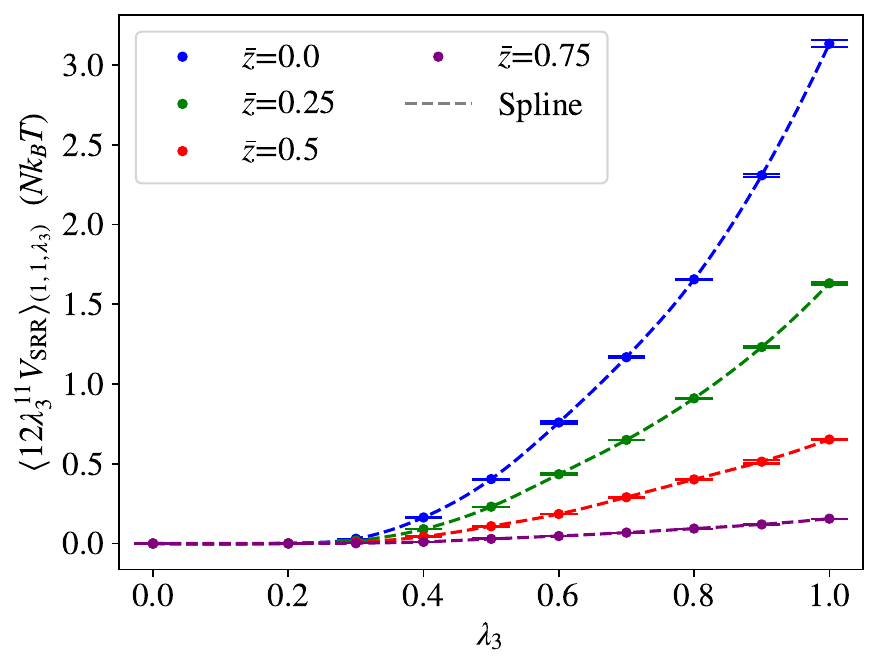}
  \caption{\label{fig:SRR_TI} Thermodynamic Integration curves  for \(r_s=2\) corresponding to the introduction of Short Range Repulsion into the BSC model. For clarity, a subset of ionisation states are displayed. The {data} was averaged over {6} independent runs, while error bars are the corresponding standard deviations.}
\end{figure}

\begin{figure}
  \centering
  \includegraphics[width=\linewidth]{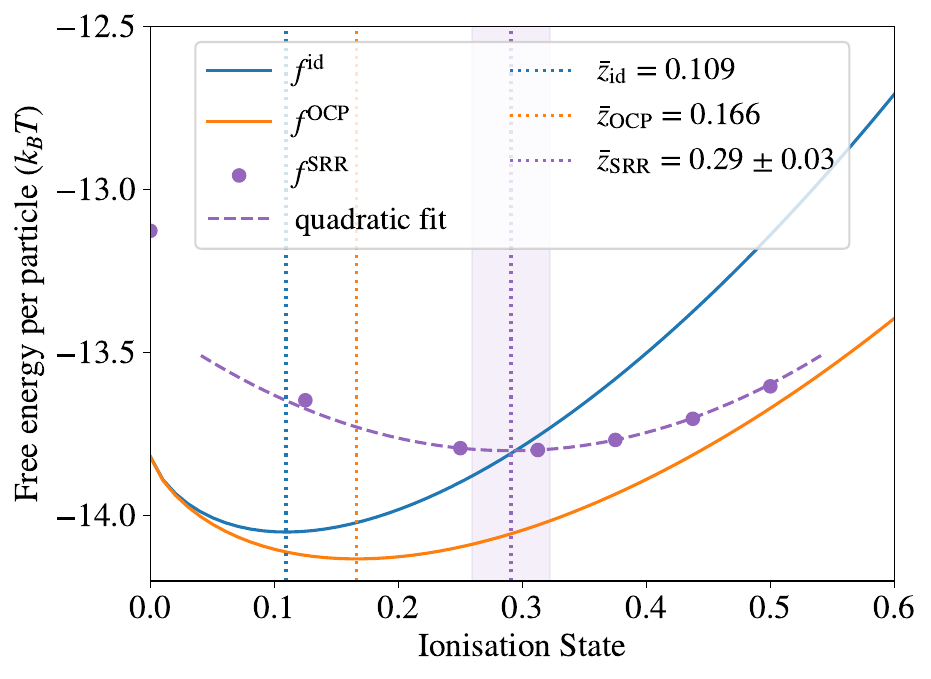}
  \caption{\label{fig:Free_Energy_SRR} SRR model free energy minimisation at \(r_s=2\). The curves with solid lines correspond to free energies of the ideal (blue, \(\vec{\lambda} = \vec{0}\)), reference (orange, \(\vec{\lambda} = (1,1,0)\)), and target (purple, \(\vec{\lambda} = \vec{1}\)) system, while the dotted lines show their minima, the uncertainty in the quadratic minimisation is also displayed in the purple shaded rectangle. The corresponding values of these minima are given in the legend.}
\end{figure}

Given these motivations, and as a demonstration of the technique, an additional SRR potential (as described in section \ref{sec:SRR_potential}) with effective radius \(r_* = 1\) has been introduced into the neutral-neutral interaction to study its effect on the predicted ionisation state. We note that such a potential could also be used to model Pauli exclusion effects between bound electrons \cite{Wunsch2009,Vorberger2013} -- these are already treated exactly in the ideal free electron gas but are not present in bound-bound or bound-free interactions. The appropriate form of the repulsive potential is unknown, but could possibly be deduced through comparison with high fidelity data, sourced from experiment or first-principles computations. Alternatively, a scheme based on the pairwise Pauli interactions from wave packet molecular dynamics  could be devised \cite{Klakow1994a, Svensson2023}. A third integration is required evaluate the influence of the harder inter-neutral core on the ionisation state. A characteristic integration curve is given in figure \ref{fig:SRR_TI} where all curves are suppressed for small \(\lambda_3\) by the factor of \((\lambda_3)^{11}\) in the derivative of the potential energy function. The free energies for this condition are displayed in figure \ref{fig:Free_Energy_SRR}. The increased atomic repulsion drives the free energy at lower ionisation states up, causing an increase in the minima. The free energy remains below the OCP free energy for higher ionisation states (not shown in figure) due to the attractive neutral-background contributions.

\subsection{Ionisation State across density}
\label{sec:ionisation_across_density}
\begin{figure}
  \centering
  \includegraphics[width=\linewidth]{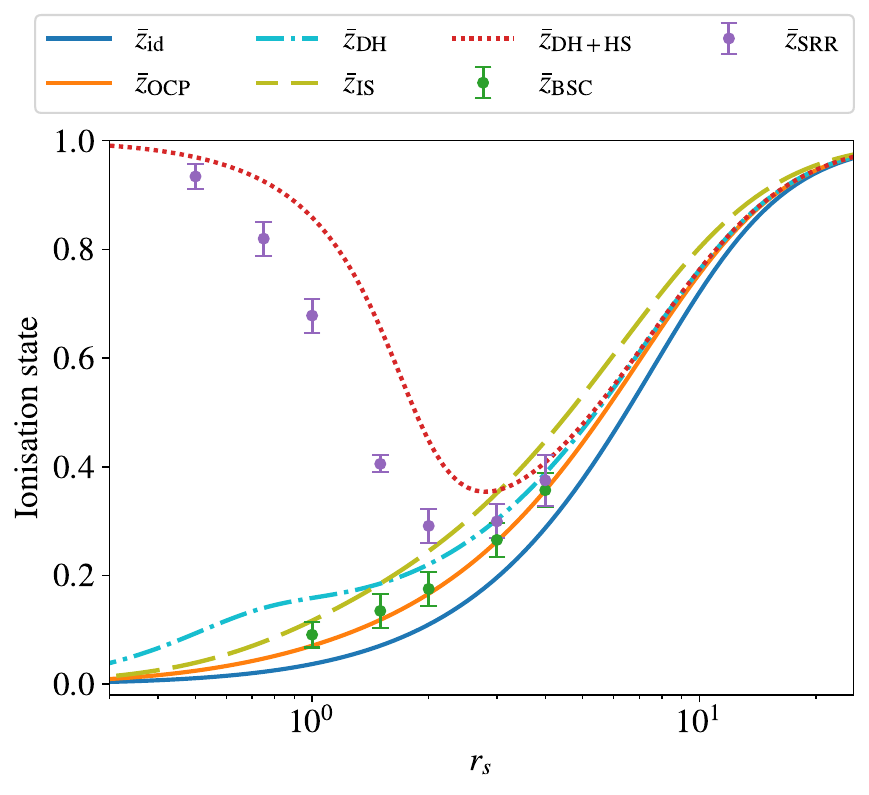}
  \caption{\label{fig:ionisation_plots} Ionisation states for models discussed in the main text plotted against \(r_s\). The BSC model (green) and SRR model (purple) correspond to results from numerically minimising free energy data. The solid curves correspond to the ideal case (blue) and OCP reference system (orange). The remaining lines relate to the free energy models defined in appendix \ref{app:ionisation_models}.}
\end{figure}

The minimisation results from the simulations performed with both the BSC model and SRR model are compiled in figure \ref{fig:ionisation_plots}. The ionisation state of the ideal hydrogen plasma decreases with increasing density.This is because the density of free particle states decreases and it becomes energetically favourable to reduce the effective number of species. Including the OCP interactions turns on long-range charged-particle interactions and increases the ionisation state, as discussed in section \ref{sec:free_min_hydrogen_model}. As \(r_s\) decreases, the coulomb coupling parameter is larger, which means the negative excess free energy contributions are greater, and the system exhibits larger binding between the ions and electron background, encouraging ionisation. At the highest densities, \(r_s \leq 1\), the free electron degeneracy contributions are dominant and the ionisation state still tends to zero. Common models for the OCP interactions are the Debye-H\"uckel and ion-sphere models which, when inserted into the minimisation procedure, give \(\bar{z}_{\rm{DH}}\) and \(\bar{z}_{\rm{IS}}\) respectively. The models are defined in appendix \ref{app:ionisation_models}. When compared with the OCP reference calculation, both models overpredict the ionisation state. The Debye-H\"uckel model is correct in the weak coupling limit, and therefore good agreement is found for \(r_s \gtrsim 5\). The ion-sphere model is more appropriate in the strongly-coupled (high density) regime, where it outperforms the Debye-H\"uckel prediction when compared against the more accurate OCP prediction. Generally, the overprediction for both models is severe, considering that the magnitude of deviation is similar to the difference between ideal and OCP curves. The deviation is directly related to both models overestimating the magnitude of the excess internal energy, which can be seen in in figure \ref{fig:OCP_data}. 

We now discuss the effect of introducing neutral interactions, starting with the BSC model introduced in \ref{sec:neutral_model}. For \(r_s \gtrsim 4\), the coulomb interactions of the neutrals have a negligible effect and the deviation in ionisation from the ideal case is purely driven by OCP interactions in the plasma. As the density increases, neutral interactions induce a small increase in the ionisation state. This is an effect of the ion-neutral and neutral-neutral repulsion effects. At the two highest densities considered: \(r_s = 0.5\) and \(r_s = 0.75\), and for high neutral fractions, the neutral-ion mixture separated into a dense ball of neutral particles surrounded by ions, hence breaking the isotropy of the fluid. This phase separation is another indication that the inter-neutral potential is too weak and these points are omitted from figure \ref{fig:ionisation_plots}. We note that at higher ionisation states, this effect is not present, and allows the BSC system to be used as an appropriate reference system when performing integration calculations with the SRR potential.

Finally, the effect of increasing the strength of collisions between neutrals has a large effect in the high density region, as expected. The pressure ionisation transition is recovered, because portions of the neutral-neutral potential that were attractive are now repulsive. This illustrates that interaction-induced ionisation is due to both long-range coulomb interactions and short-range repulsive interactions. For pressure ionisation to occur an additional repulsive interaction was artifically introduced, suggesting that effects beyond frozen-core coulomb interactions are required when modelling bound states in dense plasmas. By minimising a  Debye-H\"uckel model for the ions combined with a hard-sphere equation of state for the neutrals, the \(\bar{z}_{\rm{DH+HS}}\) curve is produced. Further details are given in appendix \ref{app:ionisation_models}. There is qualitative agreement between this model and the full SRR model. However a limited treatment of the OCP energy, neglect of the ion-neutral interaction, and replacement of the SRR potential by hard-spheres results in an overestimation of the ionisation state at high densities.

\section{Conclusion} \label{sec:conclusion}

Molecular Dynamics has a large flexibility in the choice of interparticle potential and is accurate at arbitrarily strong coupling. Meanwhile, for plasmas, the interparticle potentials are closely connected to the ionisation state. Therefore, the presented Thermodynamic Integration technique provides the possibility to extend existing models that use the chemical picture. It is also essential to determine the self-consistent ionisation state within Molecular Dynamics itself for models that involve chemical formulations such as quantum statistical potentials and wave packet molecular dynamics, which was the primary motivation for this study. Furthermore, we have investigated the ionisation state dependence on the interplay between ionic screening and neutral repulsion in hydrogen. The excluded-volume method has been used in all chemical picture based models of dense plasmas thus far to evaluate repulsive interaction effects. Given that the presented method applies to arbitrary interaction potentials, we make the first step beyond this paradigm with the bound state coulomb and short range repulsion models. 

The results presented from the one component models provide stimulus for improvement. For example, free electron interaction effects can be included either in effective ion-ion potentials \cite{Moldabekov2018b, Moldabekov2022,Blouin2021}, or in two-component models \cite{Svensson2023}. Free electrons provide additional screening and therefore would further lower the free energy at higher ionisation states \cite{Caillol2000}, causing the equilibrium ionisation state to increase. Meanwhile, multispecies systems with excited states and multiple charge states can be included within the minimisation framework. On including these developments, and leveraging the presented ionisation state calculation framework, comparison could be made against hydrogen equation of state tables \cite{Bonitz2024}.

Generally, the development of models in the chemical picture could be particularly useful in developing computational frameworks for ionisation potential depression (IPD) \cite{Bonitz2024},  given that known models falter at strong coupling \cite{Ciricosta2012}, and that IPD is non-trivial to extract from first-principles computations \cite{Gawne2023, Gawne2024,Bethkenhagen2020, Clerouin2022}. In the results presented, an IPD effect is present, and dominated by ion-ion coulomb interactions at lower densities. Furthermore, molecular dynamics provides a promising route to ionic transport coefficients, provided reliable interparticle potentials are supplied.

\begin{acknowledgments}
  We thank Thomas Campbell and Thomas Gawne for insightful discussions. We are grateful for the use of computing resources provided by STFC Scientific Computing Department’s
  SCARF cluster, where all simulations for this work were carried out. DP, PS, SMV and GG acknowledge support from AWE-NST UK via Oxford Centre for High Energy Density Science (OxCHEDs). PS acknowledges funding from the Oxford Physics Endowment for Graduates (OXPEG). S.M.V. acknowledges support from the UK EPSRC grant EP/W010097/1. The work of SMV and GG has received partial support from EPSRC and First Light Fusion under the AMPLIFI Prosperity partnership, grant no. EP/X025 373/1.
\end{acknowledgments}
  
\appendix

\section{Saha equation} \label{app:saha_equation}

The Saha equation is a common formulation used to determine the charge state distribution in partially ionised plasma \cite{Ebeling2017, Bonitz2024}. In this appendix, it is demonstrated how the free energy minimisation procedure reduces to a hydrogenic Saha equation, such that the interactions can therefore be characterised as IPD. The total free energy of the system is the sum of ideal and excess contributions:
\begin{equation}
  f = f^{\rm{id}} + f^{\rm{ex}}.
\end{equation}
Additionally, for illustrative purposes, the chemical potential of the electrons may be expressed as 
\begin{equation}
  {\mu_e} =k_B T \log \left(\frac{\bar{z}n\Lambda_e^3}{2}\right) + \Delta \mu_e,
\end{equation}
where the first term is the contribution in the non-degenerate limit, and \(\Delta \mu_e\) corrects for degeneracy. On applying the free energy minimisation, equation \eqref{eq:minimisation} and using the ideal free energy of the models, equation \eqref{eq:ideal_free}, the following relation can be obtained:
\begin{equation} \label{eq:Saha_hydrogen}
  \frac{\bar{z}^2n}{(1-\bar{z})} = \left(\frac{\Lambda_n}{\Lambda_i\Lambda_e}\right)^3 \exp\left(-\frac{I + \Delta \mu_e - \Delta I}{k_B T}\right),
\end{equation}
where \(\Delta I\) is defined as
\begin{equation} \label{eq:IPD}
  \Delta I = - k_B T \frac{\partial f^{\rm{ex}}}{\partial \bar{z}}.
\end{equation} 
In the ideal (\(\Gamma \ll 1\)) and non-degenerate (\(\theta \gg 1\)) limits: \(f^{\rm{ex}} = \Delta \mu_e =0\), and equation \eqref{eq:Saha_hydrogen} is equivalent to a common form of the hydrogenic Saha equation \cite{Ebeling2017}. Furthermore this form allows two distinct physical effects related to the ionisation state to be isolated by identifying the effective ionisation potential \cite{Bonitz2024},
\begin{equation}
  I^{\rm{eff}} = I + \Delta \mu_e - \Delta I.
\end{equation} The average energy of an electron in a Fermi gas is larger than that in a classical gas, because lower lying energy states are blocked through the Pauli principle. This means that \(\Delta \mu_e > 0\) and the effective ionisation potential is \emph{increased}. This corresponds to more energy being required to ionise a bound electron. Lastly, the presence of interactions modifies the effective ionisation potential through \(\Delta I\), which can be identified as IPD. Similar expressions have been used previously to quantify IPD \cite{Ecker1963, Griem1962}. Inspecting equation \eqref{eq:IPD}, shows that this quantity acts to reduce the ionisation potential when the excess free energy decreases with increasing ionisation. Both the OCP and SRR interactions fall into this category. As mentioned in the conclusion, our method gives direct access to the IPD and could be exploited for this purpose. We emphasise that the fidelity of the result ultimately depends upon the choice of interparticle potentials which determine the excess free energy \(f^{\rm{ex}}\).

\section{Parameterisation of OCP data} \label{app:OCP_data}

On application of equation \eqref{eq:TI_integration_2d}, the free energy due to interactions of the OCP system is found through
\begin{equation} \label{eq:free_ocp_TI}
  f^{\rm{ex}}_{\rm{OCP}}(n, T, \bar{z}) = \frac{1}{Nk_BT}\int_0^1 \langle V_{\rm{ii}} \rangle_{(\lambda_1,0)}d\lambda_1,
\end{equation}
The integral is simply equal to the excess free energy of an OCP at an effective coupling parameter \(\Gamma^{\rm{eff}} = \bar\Gamma(\bar{z}n, T)\), and equation \eqref{eq:free_ocp_TI} may be written \cite{Brush1966,Hansen1973, Caillol2010},

\begin{equation} \label{eq:free_ocp_energy}
  f^{\rm{ex}}_{\rm{OCP}}(\Gamma^{\rm{eff}}) = \bar{z}\int^{\Gamma^{\rm{eff}}}_0 \frac{u^{\rm{ex}}_{\rm{OCP}}(\Gamma)}{\Gamma} d\Gamma,
\end{equation}
where \(u^{\rm{ex}}_{\rm{OCP}} = U^{\rm{ex}}/N_i k_B T\) is the excess internal energy per particle of the OCP at a given coupling parameter, in units of \(k_B T\). We performed a parameterisation of OCP internal energy data from Monte-Carlo simulations by Caillol \cite{Caillol2010, Caillol1999}, which are well resolved in the weakly-coupled limit. A finite-size correction scheme is also applied in these works. The functional form, 
\begin{equation} \label{eq:u_ex_ocp}
  u^{\rm{ex}}_{\rm{OCP}} = \Gamma^{3/2}\left[\frac{A_1}{\sqrt{\Gamma + A_2}} + \frac{A_3}{\Gamma + 1}\right] + \frac{B_1 \Gamma^2}{\Gamma + B_2} + \frac{B_3 \Gamma^2}{\Gamma^2 + B_4},
\end{equation}
from ref. \cite{Potekhin2000} was fitted to the data which includes the asymptotic Debye-H\"uckel limit of the OCP equation of state. The fitting constants are given in table \ref{tab:coeff} and \(A_3 = -{\sqrt{3}}/{2} - {A_1}/{\sqrt{A_2}}\).

\begin{figure}
  \includegraphics[width=0.95\linewidth]{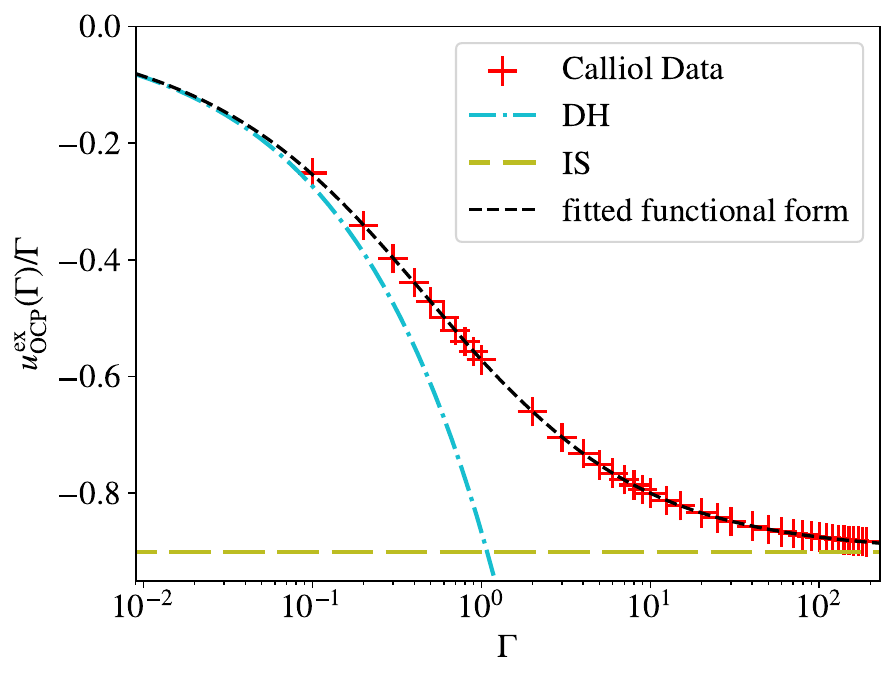}
  \caption{\label{fig:OCP_data} Thermodynamic Integration curve (black dashed line) used for the reference system in this work. The functional form from ref. \cite{Potekhin2000} is fitted to OCP data from refs \cite{Caillol1999, Caillol2010} given by the red plus symbols. The Debye-H\"uckel (DH) and ion-sphere (IS) models which are defined in appendix \ref{app:ionisation_models} are plotted for comparison.}
\end{figure}

\begin{table}
  \caption{\label{tab:coeff}%
  Coefficient values for the functional form paramterising the excess OCP energy.}
  \begin{ruledtabular}
  \begin{tabular}{cccccc}
  \(A_1\)&\(A_2\)&\(B_1\)&\(B_2\)&\(B_3\)&\(B_4\)\\
  \colrule
  -0.99787 & 0.77480 & 0.093431 & 1.5534 & 0.036253 & 4.1379\\
  \end{tabular}
  \end{ruledtabular}
  \end{table}
In figure \ref{fig:OCP_data} the excellent agreement is shown across a wide range of Coulomb coupling, the largest differences to the parameterisations given in ref. \cite{Potekhin2000} are for weak coupling, where, in this work, there is additional data for low coupling from ref. \cite{Caillol2010}. Although we note that these discrepancies are small, this regime is relevant to the overall free energy calculation because it is integrated over. The functional form allowed simple and fast evaluation of the free energy of the reference system which was found by performing the integral in equation \eqref{eq:free_ocp_energy} analytically \cite{Potekhin2000}:
\begin{equation} \label{eq:free_ocp_param}
\begin{split}
  f^{\rm{ex}}_{\rm{OCP}}(\Gamma) = & A_1 \sqrt{\Gamma (A_2 + \Gamma)} \\ &+ A_1 A_2 \log \left(-\sqrt{\Gamma/A_2}+ \sqrt{1 + \Gamma/A_2}\right) \\ &+2A_3\left[\sqrt{\Gamma} - \rm{arctan}\sqrt{\Gamma}\right] \\ & + B_1\left[\Gamma - B_2 \log \left(1 + \frac{\Gamma}{B_2}\right)\right] \\ & + \frac{B_3}{2} \log\left(1 + \frac{\Gamma^2}{B_4}\right),
\end{split}
\end{equation}
where the boundary condition \(f^{\rm{OCP}}(0) = 0\) has been applied. There is a small discrepency between line 4 in equation \eqref{eq:free_ocp_param} and the equivalent term in ref. \cite{Potekhin2000}, which we assume is due to a misprint in the latter, while the remaining terms formally agree.

\section{Finite Size Tests} \label{app:finite_size_tests}

\begin{figure}
  \includegraphics[width=0.95\linewidth]{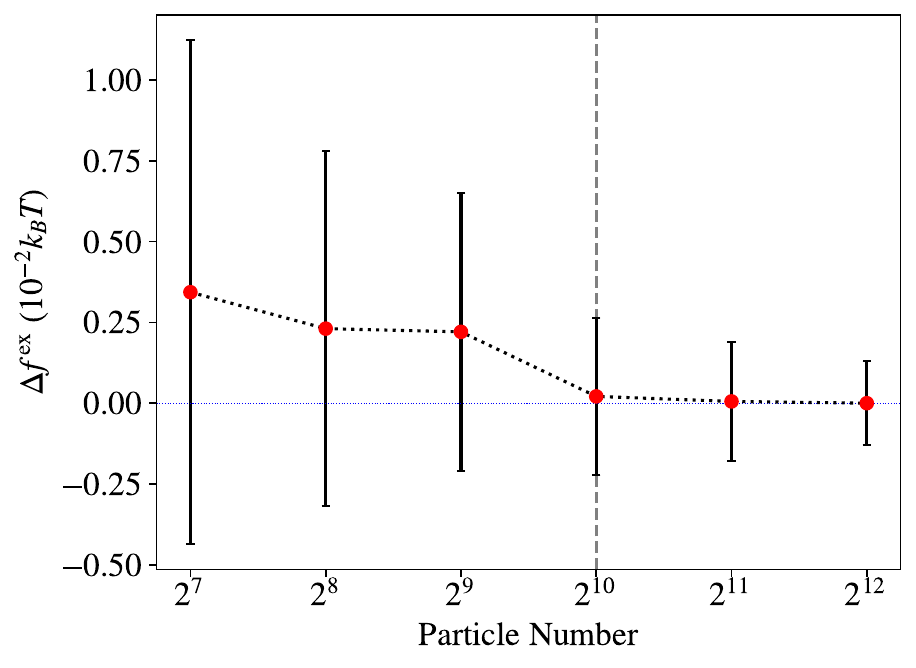}
  \caption{\label{fig:finite_size_data}  Excess free energy of the BSC model computed for different numbers of particles at a fixed density of {\(r_s = 1\)} and ionisation state of \(\bar{z}=0.5\) in units of \(10^{-2} k_B T\). The results are shown relative to the result for \(N=2^{12}\) particles: \(f^{\rm{ex}}_{\rm{BSC}}\approx-0.7124\). The dashed grey line indicates the number of particles used for the results in the main text.}
\end{figure}

Various finite-size tests were performed to check the convergence of the thermodynamic integration calculation.
In this section we present the a finite size study of the free energy calculation for an ionisation state of \(\bar{z}=0.5\) and density of \(r_s = 1\) with the BSC model. Different numbers of heavy particles ranging from \(N = 128\) to \(N=4096\) were used. To maintain a fixed density, the cubic box was rescaled accordingly with box lengths ranging from \(L \approx 8.12 \: {a}_B\) to \(L \approx 25.8 \: {a}_B\). TI curves were generated for each case with the same resolution in coupling parameter as used in figure \ref{fig:TI_example}. Each curve was then integrated to find the free energy. To calculate the statistical error in the final excess free energy, separate splines were fit to the points given by the upper and lower error bars and integrated accordingly, to give the upper and lower bounds plotted in figure \ref{fig:finite_size_data}. The statistical error is small for all cases, and decreases with system size. This is because, for larger systems, there are more trajectories to sample the integrand in equation \eqref{eq:TI_neutral}. The systematic error due to system size is also low, and the results are well converged for \(2^{10}\) particles, the number used for all results shown in the main text. Given that the TI errors are rather small, approximately \(5 \times 10^{-3} \: k_BT\) in this case, we assume the error in the full ionisation state calculation is mainly due to having a finite number of free energy data points to minimise.

\section{Ionisation models} \label{app:ionisation_models}

In section \ref{sec:ionisation_across_density}, comparison is made against simpler models which are defined in this appendix. Here they are presented as models for the excess free energy, but may be converted to an expression for IPD with equation \eqref{eq:IPD}. The first two models are concerned with the excess OCP contributions to the free energy, linked to the excess internal energy through equation \eqref{eq:free_ocp_energy}. The Debye-H\"uckel (DH) model accounts for ionic screening in the weakly coupled limit \cite{Caillol2010,Baus1980} and is given by
\begin{equation}
  f^{\rm{ex}}_{\rm{DH}} = -\bar{z}^{3/2}\frac{\Gamma^{3/2}}{\sqrt{3}}.
\end{equation}
Alternatively the Ion-Sphere (IS) model may be used, which formally provides a lower bound on the excess internal energy, as seen in figure \ref{fig:OCP_data}, and is applicable in the strongly coupled limit \cite{Baus1980, Khrapak2014}. The following expression is used,
\begin{equation} \label{eq:free_IS}
  f_{\rm{IS}}^{\rm{ex}} = -\bar{z}^{4/3}\frac{9\Gamma}{10}.
\end{equation} 
For the IS model, integration over the weakly coupled limit introduces a substantial error in the OCP free energy. Therefore alternative reference systems have been suggested in the literature by employing external data \cite{Khrapak2014}. However, to be consistent with commonly used IS models, e.g. ref. \cite{Zimmerman1980}, equation \eqref{eq:free_IS} is used in this work.

Both DH and IS models apply to the OCP subsystem and don't include neutral interactions. To estimate the effect of neutrals, the Carnahan-Stirling equation of state may be utilised \cite{Carnahan1969}. Neutrals interacting through the SRR potential with the effective radius \(r_*\) may, in a simple approximation,  be modelled as hard spheres of radius \(R = r_*\). In the model, the excess free energy due to hard-sphere interactions is
\begin{equation} \label{eq:HS}
  f^{\rm{ex}}_{\rm{HS}} = \bar{x}\frac{4\eta - 3\eta^2}{(1-\eta)^2} \quad , \quad \eta = 2\pi \bar{x} n R^3.
\end{equation}
Here \(\eta\) is the packing fraction, and, as before, \(\bar{x}\) is the neutral fraction. Finally, to estimate the effect of ion and neutral interactions simultaneously, the DH and HS free energy terms may be combined linearly, the minima of which is given by \(\bar{z}_{\rm{DH+HS}}\) in figure \ref{fig:ionisation_plots}. As mentioned in the main text, this discards the effect of ion-neutral interactions.


\bibliography{library}

\providecommand{\noopsort}[1]{}\providecommand{\singleletter}[1]{#1}%
\begin{thebibliography}{71}%
\makeatletter
\providecommand \@ifxundefined [1]{%
 \@ifx{#1\undefined}
}%
\providecommand \@ifnum [1]{%
 \ifnum #1\expandafter \@firstoftwo
 \else \expandafter \@secondoftwo
 \fi
}%
\providecommand \@ifx [1]{%
 \ifx #1\expandafter \@firstoftwo
 \else \expandafter \@secondoftwo
 \fi
}%
\providecommand \natexlab [1]{#1}%
\providecommand \enquote  [1]{``#1''}%
\providecommand \bibnamefont  [1]{#1}%
\providecommand \bibfnamefont [1]{#1}%
\providecommand \citenamefont [1]{#1}%
\providecommand \href@noop [0]{\@secondoftwo}%
\providecommand \href [0]{\begingroup \@sanitize@url \@href}%
\providecommand \@href[1]{\@@startlink{#1}\@@href}%
\providecommand \@@href[1]{\endgroup#1\@@endlink}%
\providecommand \@sanitize@url [0]{\catcode `\\12\catcode `\$12\catcode
  `\&12\catcode `\#12\catcode `\^12\catcode `\_12\catcode `\%12\relax}%
\providecommand \@@startlink[1]{}%
\providecommand \@@endlink[0]{}%
\providecommand \url  [0]{\begingroup\@sanitize@url \@url }%
\providecommand \@url [1]{\endgroup\@href {#1}{\urlprefix }}%
\providecommand \urlprefix  [0]{URL }%
\providecommand \Eprint [0]{\href }%
\providecommand \doibase [0]{https://doi.org/}%
\providecommand \selectlanguage [0]{\@gobble}%
\providecommand \bibinfo  [0]{\@secondoftwo}%
\providecommand \bibfield  [0]{\@secondoftwo}%
\providecommand \translation [1]{[#1]}%
\providecommand \BibitemOpen [0]{}%
\providecommand \bibitemStop [0]{}%
\providecommand \bibitemNoStop [0]{.\EOS\space}%
\providecommand \EOS [0]{\spacefactor3000\relax}%
\providecommand \BibitemShut  [1]{\csname bibitem#1\endcsname}%
\let\auto@bib@innerbib\@empty
\bibitem [{\citenamefont {Bonitz}\ \emph {et~al.}(2020)\citenamefont {Bonitz},
  \citenamefont {Dornheim}, \citenamefont {Moldabekov}, \citenamefont {Zhang},
  \citenamefont {Hamann}, \citenamefont {Kählert}, \citenamefont {Filinov},
  \citenamefont {Ramakrishna},\ and\ \citenamefont {Vorberger}}]{Bonitz2020}%
  \BibitemOpen
  \bibfield  {author} {\bibinfo {author} {\bibfnamefont {M.}~\bibnamefont
  {Bonitz}}, \bibinfo {author} {\bibfnamefont {T.}~\bibnamefont {Dornheim}},
  \bibinfo {author} {\bibfnamefont {Z.~A.}\ \bibnamefont {Moldabekov}},
  \bibinfo {author} {\bibfnamefont {S.}~\bibnamefont {Zhang}}, \bibinfo
  {author} {\bibfnamefont {P.}~\bibnamefont {Hamann}}, \bibinfo {author}
  {\bibfnamefont {H.}~\bibnamefont {Kählert}}, \bibinfo {author}
  {\bibfnamefont {A.}~\bibnamefont {Filinov}}, \bibinfo {author} {\bibfnamefont
  {K.}~\bibnamefont {Ramakrishna}},\ and\ \bibinfo {author} {\bibfnamefont
  {J.}~\bibnamefont {Vorberger}},\ }\bibfield  {title} {\bibinfo {title} {{Ab
  initio simulation of warm dense matter}},\ }\href
  {https://doi.org/10.1063/1.5143225} {\bibfield  {journal} {\bibinfo
  {journal} {Physics of Plasmas}\ }\textbf {\bibinfo {volume} {27}},\ \bibinfo
  {pages} {042710} (\bibinfo {year} {2020})}\BibitemShut {NoStop}%
\bibitem [{\citenamefont {Bonitz}\ \emph {et~al.}(2024)\citenamefont {Bonitz},
  \citenamefont {Vorberger}, \citenamefont {Bethkenhagen}, \citenamefont
  {Böhme}, \citenamefont {Ceperley}, \citenamefont {Filinov}, \citenamefont
  {Gawne}, \citenamefont {Graziani}, \citenamefont {Gregori}, \citenamefont
  {Hamann}, \citenamefont {Hansen}, \citenamefont {Holzmann}, \citenamefont
  {Hu}, \citenamefont {Kählert}, \citenamefont {Karasiev}, \citenamefont
  {Kleinschmidt}, \citenamefont {Kordts}, \citenamefont {Makait}, \citenamefont
  {Militzer}, \citenamefont {Moldabekov}, \citenamefont {Pierleoni},
  \citenamefont {Preising}, \citenamefont {Ramakrishna}, \citenamefont
  {Redmer}, \citenamefont {Schwalbe}, \citenamefont {Svensson},\ and\
  \citenamefont {Dornheim}}]{Bonitz2024}%
  \BibitemOpen
  \bibfield  {author} {\bibinfo {author} {\bibfnamefont {M.}~\bibnamefont
  {Bonitz}}, \bibinfo {author} {\bibfnamefont {J.}~\bibnamefont {Vorberger}},
  \bibinfo {author} {\bibfnamefont {M.}~\bibnamefont {Bethkenhagen}}, \bibinfo
  {author} {\bibfnamefont {M.}~\bibnamefont {Böhme}}, \bibinfo {author}
  {\bibfnamefont {D.}~\bibnamefont {Ceperley}}, \bibinfo {author}
  {\bibfnamefont {A.}~\bibnamefont {Filinov}}, \bibinfo {author} {\bibfnamefont
  {T.}~\bibnamefont {Gawne}}, \bibinfo {author} {\bibfnamefont
  {F.}~\bibnamefont {Graziani}}, \bibinfo {author} {\bibfnamefont
  {G.}~\bibnamefont {Gregori}}, \bibinfo {author} {\bibfnamefont
  {P.}~\bibnamefont {Hamann}}, \bibinfo {author} {\bibfnamefont
  {S.}~\bibnamefont {Hansen}}, \bibinfo {author} {\bibfnamefont
  {M.}~\bibnamefont {Holzmann}}, \bibinfo {author} {\bibfnamefont {S.~X.}\
  \bibnamefont {Hu}}, \bibinfo {author} {\bibfnamefont {H.}~\bibnamefont
  {Kählert}}, \bibinfo {author} {\bibfnamefont {V.}~\bibnamefont {Karasiev}},
  \bibinfo {author} {\bibfnamefont {U.}~\bibnamefont {Kleinschmidt}}, \bibinfo
  {author} {\bibfnamefont {L.}~\bibnamefont {Kordts}}, \bibinfo {author}
  {\bibfnamefont {C.}~\bibnamefont {Makait}}, \bibinfo {author} {\bibfnamefont
  {B.}~\bibnamefont {Militzer}}, \bibinfo {author} {\bibfnamefont
  {Z.}~\bibnamefont {Moldabekov}}, \bibinfo {author} {\bibfnamefont
  {C.}~\bibnamefont {Pierleoni}}, \bibinfo {author} {\bibfnamefont
  {M.}~\bibnamefont {Preising}}, \bibinfo {author} {\bibfnamefont
  {K.}~\bibnamefont {Ramakrishna}}, \bibinfo {author} {\bibfnamefont
  {R.}~\bibnamefont {Redmer}}, \bibinfo {author} {\bibfnamefont
  {S.}~\bibnamefont {Schwalbe}}, \bibinfo {author} {\bibfnamefont
  {P.}~\bibnamefont {Svensson}},\ and\ \bibinfo {author} {\bibfnamefont
  {T.}~\bibnamefont {Dornheim}},\ }\href {https://arxiv.org/abs/2405.10627}
  {\bibinfo {title} {First principles simulations of dense hydrogen}} (\bibinfo
  {year} {2024}),\ \Eprint {https://arxiv.org/abs/2405.10627} {arXiv:2405.10627
  [physics.comp-ph]} \BibitemShut {NoStop}%
\bibitem [{\citenamefont {Dornheim}\ \emph {et~al.}(2018)\citenamefont
  {Dornheim}, \citenamefont {Groth},\ and\ \citenamefont
  {Bonitz}}]{Dornheim2018}%
  \BibitemOpen
  \bibfield  {author} {\bibinfo {author} {\bibfnamefont {T.}~\bibnamefont
  {Dornheim}}, \bibinfo {author} {\bibfnamefont {S.}~\bibnamefont {Groth}},\
  and\ \bibinfo {author} {\bibfnamefont {M.}~\bibnamefont {Bonitz}},\
  }\bibfield  {title} {\bibinfo {title} {The uniform electron gas at warm dense
  matter conditions},\ }\href
  {https://doi.org/https://doi.org/10.1016/j.physrep.2018.04.001} {\bibfield
  {journal} {\bibinfo  {journal} {Physics Reports}\ }\textbf {\bibinfo {volume}
  {744}},\ \bibinfo {pages} {1} (\bibinfo {year} {2018})}\BibitemShut {NoStop}%
\bibitem [{\citenamefont {Filinov}\ and\ \citenamefont
  {Bonitz}(2023)}]{Filinov2023}%
  \BibitemOpen
  \bibfield  {author} {\bibinfo {author} {\bibfnamefont {A.~V.}\ \bibnamefont
  {Filinov}}\ and\ \bibinfo {author} {\bibfnamefont {M.}~\bibnamefont
  {Bonitz}},\ }\bibfield  {title} {\bibinfo {title} {Equation of state of
  partially ionized hydrogen and deuterium plasma revisited},\ }\href
  {https://doi.org/10.1103/PHYSREVE.108.055212/FIGURES/19/MEDIUM} {\bibfield
  {journal} {\bibinfo  {journal} {Physical Review E}\ }\textbf {\bibinfo
  {volume} {108}},\ \bibinfo {pages} {055212} (\bibinfo {year}
  {2023})}\BibitemShut {NoStop}%
\bibitem [{\citenamefont {Marx}\ and\ \citenamefont
  {Hutter}(2009)}]{Hutter2009}%
  \BibitemOpen
  \bibfield  {author} {\bibinfo {author} {\bibfnamefont {D.}~\bibnamefont
  {Marx}}\ and\ \bibinfo {author} {\bibfnamefont {J.}~\bibnamefont {Hutter}},\
  }\href@noop {} {\emph {\bibinfo {title} {Ab Initio Molecular Dynamics: Basic
  Theory and Advanced Methods}}}\ (\bibinfo  {publisher} {Cambridge University
  Press},\ \bibinfo {year} {2009})\BibitemShut {NoStop}%
\bibitem [{\citenamefont {Moldabekov}\ \emph {et~al.}(2018)\citenamefont
  {Moldabekov}, \citenamefont {Groth}, \citenamefont {Dornheim}, \citenamefont
  {K\"ahlert}, \citenamefont {Bonitz},\ and\ \citenamefont
  {Ramazanov}}]{Moldabekov2018b}%
  \BibitemOpen
  \bibfield  {author} {\bibinfo {author} {\bibfnamefont {Z.~A.}\ \bibnamefont
  {Moldabekov}}, \bibinfo {author} {\bibfnamefont {S.}~\bibnamefont {Groth}},
  \bibinfo {author} {\bibfnamefont {T.}~\bibnamefont {Dornheim}}, \bibinfo
  {author} {\bibfnamefont {H.}~\bibnamefont {K\"ahlert}}, \bibinfo {author}
  {\bibfnamefont {M.}~\bibnamefont {Bonitz}},\ and\ \bibinfo {author}
  {\bibfnamefont {T.~S.}\ \bibnamefont {Ramazanov}},\ }\bibfield  {title}
  {\bibinfo {title} {Structural characteristics of strongly coupled ions in a
  dense quantum plasma},\ }\href {https://doi.org/10.1103/PhysRevE.98.023207}
  {\bibfield  {journal} {\bibinfo  {journal} {Phys. Rev. E}\ }\textbf {\bibinfo
  {volume} {98}},\ \bibinfo {pages} {023207} (\bibinfo {year}
  {2018})}\BibitemShut {NoStop}%
\bibitem [{\citenamefont {Moldabekov}\ \emph {et~al.}(2019)\citenamefont
  {Moldabekov}, \citenamefont {K\"ahlert}, \citenamefont {Dornheim},
  \citenamefont {Groth}, \citenamefont {Bonitz},\ and\ \citenamefont
  {Ramazanov}}]{Moldabekov2019}%
  \BibitemOpen
  \bibfield  {author} {\bibinfo {author} {\bibfnamefont {Z.~A.}\ \bibnamefont
  {Moldabekov}}, \bibinfo {author} {\bibfnamefont {H.}~\bibnamefont
  {K\"ahlert}}, \bibinfo {author} {\bibfnamefont {T.}~\bibnamefont {Dornheim}},
  \bibinfo {author} {\bibfnamefont {S.}~\bibnamefont {Groth}}, \bibinfo
  {author} {\bibfnamefont {M.}~\bibnamefont {Bonitz}},\ and\ \bibinfo {author}
  {\bibfnamefont {T.~S.}\ \bibnamefont {Ramazanov}},\ }\bibfield  {title}
  {\bibinfo {title} {Dynamical structure factor of strongly coupled ions in a
  dense quantum plasma},\ }\href {https://doi.org/10.1103/PhysRevE.99.053203}
  {\bibfield  {journal} {\bibinfo  {journal} {Phys. Rev. E}\ }\textbf {\bibinfo
  {volume} {99}},\ \bibinfo {pages} {053203} (\bibinfo {year}
  {2019})}\BibitemShut {NoStop}%
\bibitem [{\citenamefont {Blouin}\ and\ \citenamefont
  {Daligault}(2021)}]{Blouin2021}%
  \BibitemOpen
  \bibfield  {author} {\bibinfo {author} {\bibfnamefont {S.}~\bibnamefont
  {Blouin}}\ and\ \bibinfo {author} {\bibfnamefont {J.}~\bibnamefont
  {Daligault}},\ }\bibfield  {title} {\bibinfo {title} {Direct evaluation of
  the phase diagrams of dense multicomponent plasmas by integration of the
  clapeyron equations},\ }\href {https://doi.org/10.1103/PhysRevE.103.043204}
  {\bibfield  {journal} {\bibinfo  {journal} {Phys. Rev. E}\ }\textbf {\bibinfo
  {volume} {103}},\ \bibinfo {pages} {043204} (\bibinfo {year}
  {2021})}\BibitemShut {NoStop}%
\bibitem [{\citenamefont {Brush}\ \emph {et~al.}(1966)\citenamefont {Brush},
  \citenamefont {Sahlin},\ and\ \citenamefont {Teller}}]{Brush1966}%
  \BibitemOpen
  \bibfield  {author} {\bibinfo {author} {\bibfnamefont {S.~G.}\ \bibnamefont
  {Brush}}, \bibinfo {author} {\bibfnamefont {H.~L.}\ \bibnamefont {Sahlin}},\
  and\ \bibinfo {author} {\bibfnamefont {E.}~\bibnamefont {Teller}},\
  }\bibfield  {title} {\bibinfo {title} {Monte carlo study of a one‐component
  plasma. {I}},\ }\href {https://doi.org/10.1063/1.1727895} {\bibfield
  {journal} {\bibinfo  {journal} {The Journal of Chemical Physics}\ }\textbf
  {\bibinfo {volume} {45}},\ \bibinfo {pages} {2102} (\bibinfo {year}
  {1966})}\BibitemShut {NoStop}%
\bibitem [{\citenamefont {Hansen}(1973)}]{Hansen1973}%
  \BibitemOpen
  \bibfield  {author} {\bibinfo {author} {\bibfnamefont {J.~P.}\ \bibnamefont
  {Hansen}},\ }\bibfield  {title} {\bibinfo {title} {Statistical mechanics of
  dense ionized matter. {I}. equilibrium properties of the classical
  one-component plasma},\ }\href {https://doi.org/10.1103/PhysRevA.8.3096}
  {\bibfield  {journal} {\bibinfo  {journal} {Phys. Rev. A}\ }\textbf {\bibinfo
  {volume} {8}},\ \bibinfo {pages} {3096} (\bibinfo {year} {1973})}\BibitemShut
  {NoStop}%
\bibitem [{\citenamefont {Caillol}(1999{\natexlab{a}})}]{Caillol1999}%
  \BibitemOpen
  \bibfield  {author} {\bibinfo {author} {\bibfnamefont {J.~M.}\ \bibnamefont
  {Caillol}},\ }\bibfield  {title} {\bibinfo {title} {Thermodynamic limit of
  the excess internal energy of the fluid phase of a one-component plasma: A
  monte carlo study},\ }\href {https://doi.org/10.1063/1.479965} {\bibfield
  {journal} {\bibinfo  {journal} {Journal of Chemical Physics}\ }\textbf
  {\bibinfo {volume} {111}},\ \bibinfo {pages} {6538} (\bibinfo {year}
  {1999}{\natexlab{a}})}\BibitemShut {NoStop}%
\bibitem [{\citenamefont {Caillol}\ and\ \citenamefont
  {Gilles}(2010)}]{Caillol2010}%
  \BibitemOpen
  \bibfield  {author} {\bibinfo {author} {\bibfnamefont {J.-M.}\ \bibnamefont
  {Caillol}}\ and\ \bibinfo {author} {\bibfnamefont {D.}~\bibnamefont
  {Gilles}},\ }\bibfield  {title} {\bibinfo {title} {An accurate equation of
  state for the one-component plasma in the low coupling regime},\ }\href
  {https://doi.org/10.1088/1751-8113/43/10/105501} {\bibfield  {journal}
  {\bibinfo  {journal} {Journal of Physics A: Mathematical and Theoretical}\
  }\textbf {\bibinfo {volume} {43}},\ \bibinfo {pages} {105501} (\bibinfo
  {year} {2010})}\BibitemShut {NoStop}%
\bibitem [{\citenamefont {Onegin}\ \emph {et~al.}(2024)\citenamefont {Onegin},
  \citenamefont {Demyanov},\ and\ \citenamefont {Levashov}}]{Onegin2024}%
  \BibitemOpen
  \bibfield  {author} {\bibinfo {author} {\bibfnamefont {A.~S.}\ \bibnamefont
  {Onegin}}, \bibinfo {author} {\bibfnamefont {G.~S.}\ \bibnamefont
  {Demyanov}},\ and\ \bibinfo {author} {\bibfnamefont {P.~R.}\ \bibnamefont
  {Levashov}},\ }\bibfield  {title} {\bibinfo {title} {Pressure of coulomb
  systems with volume-dependent long-range potentials},\ }\href
  {https://doi.org/10.1088/1751-8121/ad40e5} {\bibfield  {journal} {\bibinfo
  {journal} {Journal of Physics A: Mathematical and Theoretical}\ }\textbf
  {\bibinfo {volume} {57}},\ \bibinfo {pages} {205002} (\bibinfo {year}
  {2024})}\BibitemShut {NoStop}%
\bibitem [{\citenamefont {Khrapak}\ and\ \citenamefont
  {Khrapak}(2014)}]{Khrapak2014}%
  \BibitemOpen
  \bibfield  {author} {\bibinfo {author} {\bibfnamefont {S.~A.}\ \bibnamefont
  {Khrapak}}\ and\ \bibinfo {author} {\bibfnamefont {A.~G.}\ \bibnamefont
  {Khrapak}},\ }\bibfield  {title} {\bibinfo {title} {{Simple thermodynamics of
  strongly coupled one-component-plasma in two and three dimensions}},\ }\href
  {https://doi.org/10.1063/1.4897386} {\bibfield  {journal} {\bibinfo
  {journal} {Physics of Plasmas}\ }\textbf {\bibinfo {volume} {21}},\ \bibinfo
  {pages} {104505} (\bibinfo {year} {2014})}\BibitemShut {NoStop}%
\bibitem [{\citenamefont {Baus}\ and\ \citenamefont {Hansen}(1980)}]{Baus1980}%
  \BibitemOpen
  \bibfield  {author} {\bibinfo {author} {\bibfnamefont {M.}~\bibnamefont
  {Baus}}\ and\ \bibinfo {author} {\bibfnamefont {J.-P.}\ \bibnamefont
  {Hansen}},\ }\bibfield  {title} {\bibinfo {title} {Statistical mechanics of
  simple coulomb systems},\ }\href
  {https://doi.org/https://doi.org/10.1016/0370-1573(80)90022-8} {\bibfield
  {journal} {\bibinfo  {journal} {Physics Reports}\ }\textbf {\bibinfo {volume}
  {59}},\ \bibinfo {pages} {1} (\bibinfo {year} {1980})}\BibitemShut {NoStop}%
\bibitem [{\citenamefont {Demyanov}\ and\ \citenamefont
  {Levashov}(2022)}]{Demyanov2022}%
  \BibitemOpen
  \bibfield  {author} {\bibinfo {author} {\bibfnamefont {G.~S.}\ \bibnamefont
  {Demyanov}}\ and\ \bibinfo {author} {\bibfnamefont {P.~R.}\ \bibnamefont
  {Levashov}},\ }\bibfield  {title} {\bibinfo {title} {One-component plasma of
  a million particles via angular-averaged ewald potential: A monte carlo
  study},\ }\href {https://doi.org/10.1103/PhysRevE.106.015204} {\bibfield
  {journal} {\bibinfo  {journal} {Phys. Rev. E}\ }\textbf {\bibinfo {volume}
  {106}},\ \bibinfo {pages} {015204} (\bibinfo {year} {2022})}\BibitemShut
  {NoStop}%
\bibitem [{\citenamefont {K\"ahlert}(2020)}]{Kahlert2020}%
  \BibitemOpen
  \bibfield  {author} {\bibinfo {author} {\bibfnamefont {H.}~\bibnamefont
  {K\"ahlert}},\ }\bibfield  {title} {\bibinfo {title} {Thermodynamic and
  transport coefficients from the dynamic structure factor of yukawa liquids},\
  }\href {https://doi.org/10.1103/PhysRevResearch.2.033287} {\bibfield
  {journal} {\bibinfo  {journal} {Phys. Rev. Res.}\ }\textbf {\bibinfo {volume}
  {2}},\ \bibinfo {pages} {033287} (\bibinfo {year} {2020})}\BibitemShut
  {NoStop}%
\bibitem [{\citenamefont {Mithen}\ \emph {et~al.}(2011)\citenamefont {Mithen},
  \citenamefont {Daligault},\ and\ \citenamefont {Gregori}}]{Mithen2011}%
  \BibitemOpen
  \bibfield  {author} {\bibinfo {author} {\bibfnamefont {J.~P.}\ \bibnamefont
  {Mithen}}, \bibinfo {author} {\bibfnamefont {J.}~\bibnamefont {Daligault}},\
  and\ \bibinfo {author} {\bibfnamefont {G.}~\bibnamefont {Gregori}},\
  }\bibfield  {title} {\bibinfo {title} {Extent of validity of the hydrodynamic
  description of ions in dense plasmas},\ }\href
  {https://doi.org/10.1103/PhysRevE.83.015401} {\bibfield  {journal} {\bibinfo
  {journal} {Phys. Rev. E}\ }\textbf {\bibinfo {volume} {83}},\ \bibinfo
  {pages} {015401(R)} (\bibinfo {year} {2011})}\BibitemShut {NoStop}%
\bibitem [{\citenamefont {Klakow}\ \emph
  {et~al.}(1994{\natexlab{a}})\citenamefont {Klakow}, \citenamefont
  {Toepffer},\ and\ \citenamefont {Reinhard}}]{Klakow1994a}%
  \BibitemOpen
  \bibfield  {author} {\bibinfo {author} {\bibfnamefont {D.}~\bibnamefont
  {Klakow}}, \bibinfo {author} {\bibfnamefont {C.}~\bibnamefont {Toepffer}},\
  and\ \bibinfo {author} {\bibfnamefont {P.-G.}\ \bibnamefont {Reinhard}},\
  }\bibfield  {title} {\bibinfo {title} {Hydrogen under extreme conditions},\
  }\href {https://doi.org/https://doi.org/10.1016/0375-9601(94)91015-4}
  {\bibfield  {journal} {\bibinfo  {journal} {Physics Letters A}\ }\textbf
  {\bibinfo {volume} {192}},\ \bibinfo {pages} {55} (\bibinfo {year}
  {1994}{\natexlab{a}})}\BibitemShut {NoStop}%
\bibitem [{\citenamefont {Klakow}\ \emph
  {et~al.}(1994{\natexlab{b}})\citenamefont {Klakow}, \citenamefont
  {Toepffer},\ and\ \citenamefont {Reinhard}}]{Klakow1994b}%
  \BibitemOpen
  \bibfield  {author} {\bibinfo {author} {\bibfnamefont {D.}~\bibnamefont
  {Klakow}}, \bibinfo {author} {\bibfnamefont {C.}~\bibnamefont {Toepffer}},\
  and\ \bibinfo {author} {\bibfnamefont {P.}~\bibnamefont {Reinhard}},\
  }\bibfield  {title} {\bibinfo {title} {{Semiclassical molecular dynamics for
  strongly coupled Coulomb systems}},\ }\href
  {https://doi.org/10.1063/1.467889} {\bibfield  {journal} {\bibinfo  {journal}
  {The Journal of Chemical Physics}\ }\textbf {\bibinfo {volume} {101}},\
  \bibinfo {pages} {10766} (\bibinfo {year} {1994}{\natexlab{b}})}\BibitemShut
  {NoStop}%
\bibitem [{\citenamefont {Ebeling}\ and\ \citenamefont
  {Militzer}(1997)}]{Ebeling1997}%
  \BibitemOpen
  \bibfield  {author} {\bibinfo {author} {\bibfnamefont {W.}~\bibnamefont
  {Ebeling}}\ and\ \bibinfo {author} {\bibfnamefont {B.}~\bibnamefont
  {Militzer}},\ }\bibfield  {title} {\bibinfo {title} {Quantum molecular
  dynamics of partially ionized plasmas},\ }\href
  {https://doi.org/10.1016/S0375-9601(96)00948-6} {\bibfield  {journal}
  {\bibinfo  {journal} {Physics Letters A}\ }\textbf {\bibinfo {volume}
  {226}},\ \bibinfo {pages} {298} (\bibinfo {year} {1997})}\BibitemShut
  {NoStop}%
\bibitem [{\citenamefont {Knaup}\ \emph {et~al.}(1999)\citenamefont {Knaup},
  \citenamefont {Reinhard},\ and\ \citenamefont {Toepffer}}]{Knaup1999}%
  \BibitemOpen
  \bibfield  {author} {\bibinfo {author} {\bibfnamefont {M.}~\bibnamefont
  {Knaup}}, \bibinfo {author} {\bibfnamefont {P.-G.}\ \bibnamefont
  {Reinhard}},\ and\ \bibinfo {author} {\bibfnamefont {C.}~\bibnamefont
  {Toepffer}},\ }\bibfield  {title} {\bibinfo {title} {Wave packet molecular
  dynamics simulations of hydrogen near the transition to a metallic fluid},\
  }\href {https://doi.org/https://doi.org/10.1002/ctpp.2150390114} {\bibfield
  {journal} {\bibinfo  {journal} {Contributions to Plasma Physics}\ }\textbf
  {\bibinfo {volume} {39}},\ \bibinfo {pages} {57} (\bibinfo {year}
  {1999})}\BibitemShut {NoStop}%
\bibitem [{\citenamefont {Lavrinenko}\ \emph {et~al.}(2021)\citenamefont
  {Lavrinenko}, \citenamefont {Levashov}, \citenamefont {Minakov},
  \citenamefont {Morozov},\ and\ \citenamefont {Valuev}}]{Lavrinenko2021}%
  \BibitemOpen
  \bibfield  {author} {\bibinfo {author} {\bibfnamefont {Y.}~\bibnamefont
  {Lavrinenko}}, \bibinfo {author} {\bibfnamefont {P.~R.}\ \bibnamefont
  {Levashov}}, \bibinfo {author} {\bibfnamefont {D.~V.}\ \bibnamefont
  {Minakov}}, \bibinfo {author} {\bibfnamefont {I.~V.}\ \bibnamefont
  {Morozov}},\ and\ \bibinfo {author} {\bibfnamefont {I.~A.}\ \bibnamefont
  {Valuev}},\ }\bibfield  {title} {\bibinfo {title} {Equilibrium properties of
  warm dense deuterium calculated by the wave packet molecular dynamics and
  density functional theory method},\ }\href
  {https://doi.org/10.1103/PhysRevE.104.045304} {\bibfield  {journal} {\bibinfo
   {journal} {Phys. Rev. E}\ }\textbf {\bibinfo {volume} {104}},\ \bibinfo
  {pages} {045304} (\bibinfo {year} {2021})}\BibitemShut {NoStop}%
\bibitem [{\citenamefont {Davis}\ \emph {et~al.}(2020)\citenamefont {Davis},
  \citenamefont {Angermeier}, \citenamefont {Hermsmeier},\ and\ \citenamefont
  {White}}]{Davis2020}%
  \BibitemOpen
  \bibfield  {author} {\bibinfo {author} {\bibfnamefont {R.~A.}\ \bibnamefont
  {Davis}}, \bibinfo {author} {\bibfnamefont {W.~A.}\ \bibnamefont
  {Angermeier}}, \bibinfo {author} {\bibfnamefont {R.~K.~T.}\ \bibnamefont
  {Hermsmeier}},\ and\ \bibinfo {author} {\bibfnamefont {T.~G.}\ \bibnamefont
  {White}},\ }\bibfield  {title} {\bibinfo {title} {Ion modes in dense ionized
  plasmas through nonadiabatic molecular dynamics},\ }\href
  {https://doi.org/10.1103/PhysRevResearch.2.043139} {\bibfield  {journal}
  {\bibinfo  {journal} {Phys. Rev. Res.}\ }\textbf {\bibinfo {volume} {2}},\
  \bibinfo {pages} {043139} (\bibinfo {year} {2020})}\BibitemShut {NoStop}%
\bibitem [{\citenamefont {Svensson}\ \emph {et~al.}(2023)\citenamefont
  {Svensson}, \citenamefont {Campbell}, \citenamefont {Graziani}, \citenamefont
  {Moldabekov}, \citenamefont {Lyu}, \citenamefont {Batista}, \citenamefont
  {Richardson}, \citenamefont {Vinko},\ and\ \citenamefont
  {Gregori}}]{Svensson2023}%
  \BibitemOpen
  \bibfield  {author} {\bibinfo {author} {\bibfnamefont {P.}~\bibnamefont
  {Svensson}}, \bibinfo {author} {\bibfnamefont {T.}~\bibnamefont {Campbell}},
  \bibinfo {author} {\bibfnamefont {F.}~\bibnamefont {Graziani}}, \bibinfo
  {author} {\bibfnamefont {Z.}~\bibnamefont {Moldabekov}}, \bibinfo {author}
  {\bibfnamefont {N.}~\bibnamefont {Lyu}}, \bibinfo {author} {\bibfnamefont
  {V.~S.}\ \bibnamefont {Batista}}, \bibinfo {author} {\bibfnamefont
  {S.}~\bibnamefont {Richardson}}, \bibinfo {author} {\bibfnamefont {S.~M.}\
  \bibnamefont {Vinko}},\ and\ \bibinfo {author} {\bibfnamefont
  {G.}~\bibnamefont {Gregori}},\ }\bibfield  {title} {\bibinfo {title}
  {Development of a new quantum trajectory molecular dynamics framework},\
  }\bibfield  {journal} {\bibinfo  {journal} {Philosophical Transactions of the
  Royal Society A}\ }\textbf {\bibinfo {volume} {381}},\ \href
  {https://doi.org/10.1098/RSTA.2022.0325} {10.1098/RSTA.2022.0325} (\bibinfo
  {year} {2023})\BibitemShut {NoStop}%
\bibitem [{\citenamefont {Hansen}\ and\ \citenamefont
  {McDonald}(1978)}]{Hansen1978}%
  \BibitemOpen
  \bibfield  {author} {\bibinfo {author} {\bibfnamefont {J.~P.}\ \bibnamefont
  {Hansen}}\ and\ \bibinfo {author} {\bibfnamefont {I.~R.}\ \bibnamefont
  {McDonald}},\ }\bibfield  {title} {\bibinfo {title} {Microscopic simulation
  of a hydrogen plasma},\ }\href {https://doi.org/10.1103/PhysRevLett.41.1379}
  {\bibfield  {journal} {\bibinfo  {journal} {Phys. Rev. Lett.}\ }\textbf
  {\bibinfo {volume} {41}},\ \bibinfo {pages} {1379} (\bibinfo {year}
  {1978})}\BibitemShut {NoStop}%
\bibitem [{\citenamefont {Filinov}\ \emph {et~al.}(2004)\citenamefont
  {Filinov}, \citenamefont {Golubnychiy}, \citenamefont {Bonitz}, \citenamefont
  {Ebeling},\ and\ \citenamefont {Dufty}}]{Filinov2004}%
  \BibitemOpen
  \bibfield  {author} {\bibinfo {author} {\bibfnamefont {A.~V.}\ \bibnamefont
  {Filinov}}, \bibinfo {author} {\bibfnamefont {V.~O.}\ \bibnamefont
  {Golubnychiy}}, \bibinfo {author} {\bibfnamefont {M.}~\bibnamefont {Bonitz}},
  \bibinfo {author} {\bibfnamefont {W.}~\bibnamefont {Ebeling}},\ and\ \bibinfo
  {author} {\bibfnamefont {J.~W.}\ \bibnamefont {Dufty}},\ }\bibfield  {title}
  {\bibinfo {title} {Temperature-dependent quantum pair potentials and their
  application to dense partially ionized hydrogen plasmas},\ }\href
  {https://doi.org/10.1103/PhysRevE.70.046411} {\bibfield  {journal} {\bibinfo
  {journal} {Phys. Rev. E}\ }\textbf {\bibinfo {volume} {70}},\ \bibinfo
  {pages} {046411} (\bibinfo {year} {2004})}\BibitemShut {NoStop}%
\bibitem [{\citenamefont {Ebeling}\ \emph {et~al.}(2006)\citenamefont
  {Ebeling}, \citenamefont {Filinov}, \citenamefont {Bonitz}, \citenamefont
  {Filinov},\ and\ \citenamefont {Pohl}}]{Ebeling2006}%
  \BibitemOpen
  \bibfield  {author} {\bibinfo {author} {\bibfnamefont {W.}~\bibnamefont
  {Ebeling}}, \bibinfo {author} {\bibfnamefont {A.}~\bibnamefont {Filinov}},
  \bibinfo {author} {\bibfnamefont {M.}~\bibnamefont {Bonitz}}, \bibinfo
  {author} {\bibfnamefont {V.}~\bibnamefont {Filinov}},\ and\ \bibinfo {author}
  {\bibfnamefont {T.}~\bibnamefont {Pohl}},\ }\bibfield  {title} {\bibinfo
  {title} {The method of effective potentials in the quantum-statistical theory
  of plasmas},\ }\href {https://doi.org/10.1088/0305-4470/39/17/S01} {\bibfield
   {journal} {\bibinfo  {journal} {Journal of Physics A: Mathematical and
  General}\ }\textbf {\bibinfo {volume} {39}},\ \bibinfo {pages} {4309}
  (\bibinfo {year} {2006})}\BibitemShut {NoStop}%
\bibitem [{\citenamefont {Calisti}\ \emph {et~al.}(2024)\citenamefont
  {Calisti}, \citenamefont {Ferri}, \citenamefont {Mossé},\ and\ \citenamefont
  {Talin}}]{Calisti2024}%
  \BibitemOpen
  \bibfield  {author} {\bibinfo {author} {\bibfnamefont {A.}~\bibnamefont
  {Calisti}}, \bibinfo {author} {\bibfnamefont {S.}~\bibnamefont {Ferri}},
  \bibinfo {author} {\bibfnamefont {C.}~\bibnamefont {Mossé}},\ and\ \bibinfo
  {author} {\bibfnamefont {B.}~\bibnamefont {Talin}},\ }\bibfield  {title}
  {\bibinfo {title} {Classical molecular dynamic codes for hot dense plasmas:
  The bingo code suite},\ }\href
  {https://doi.org/https://doi.org/10.1016/j.hedp.2024.101084} {\bibfield
  {journal} {\bibinfo  {journal} {High Energy Density Physics}\ }\textbf
  {\bibinfo {volume} {50}},\ \bibinfo {pages} {101084} (\bibinfo {year}
  {2024})}\BibitemShut {NoStop}%
\bibitem [{\citenamefont {Larder}\ \emph {et~al.}(2019)\citenamefont {Larder},
  \citenamefont {Gericke}, \citenamefont {Richardson}, \citenamefont {Mabey},
  \citenamefont {White},\ and\ \citenamefont {Gregori}}]{Larder2019}%
  \BibitemOpen
  \bibfield  {author} {\bibinfo {author} {\bibfnamefont {B.}~\bibnamefont
  {Larder}}, \bibinfo {author} {\bibfnamefont {D.~O.}\ \bibnamefont {Gericke}},
  \bibinfo {author} {\bibfnamefont {S.}~\bibnamefont {Richardson}}, \bibinfo
  {author} {\bibfnamefont {P.}~\bibnamefont {Mabey}}, \bibinfo {author}
  {\bibfnamefont {T.~G.}\ \bibnamefont {White}},\ and\ \bibinfo {author}
  {\bibfnamefont {G.}~\bibnamefont {Gregori}},\ }\bibfield  {title} {\bibinfo
  {title} {Fast nonadiabatic dynamics of many-body quantum systems},\ }\href
  {https://doi.org/10.1126/sciadv.aaw1634} {\bibfield  {journal} {\bibinfo
  {journal} {Science Advances}\ }\textbf {\bibinfo {volume} {5}},\ \bibinfo
  {pages} {eaaw1634} (\bibinfo {year} {2019})}\BibitemShut {NoStop}%
\bibitem [{\citenamefont {Campbell}\ \emph {et~al.}(2024)\citenamefont
  {Campbell}, \citenamefont {Svensson}, \citenamefont {Larder}, \citenamefont
  {Plummer}, \citenamefont {Vinko},\ and\ \citenamefont
  {Gregori}}]{Campbell2024}%
  \BibitemOpen
  \bibfield  {author} {\bibinfo {author} {\bibfnamefont {T.}~\bibnamefont
  {Campbell}}, \bibinfo {author} {\bibfnamefont {P.}~\bibnamefont {Svensson}},
  \bibinfo {author} {\bibfnamefont {B.}~\bibnamefont {Larder}}, \bibinfo
  {author} {\bibfnamefont {D.}~\bibnamefont {Plummer}}, \bibinfo {author}
  {\bibfnamefont {S.~M.}\ \bibnamefont {Vinko}},\ and\ \bibinfo {author}
  {\bibfnamefont {G.}~\bibnamefont {Gregori}},\ }\href
  {https://arxiv.org/abs/2408.03693} {\bibinfo {title} {A molecular dynamics
  framework coupled with smoothed particle hydrodynamics for quantum plasma
  simulations}} (\bibinfo {year} {2024}),\ \Eprint
  {https://arxiv.org/abs/2408.03693} {arXiv:2408.03693 [physics.plasm-ph]}
  \BibitemShut {NoStop}%
\bibitem [{\citenamefont {Svensson}\ \emph
  {et~al.}(2024{\natexlab{a}})\citenamefont {Svensson}, \citenamefont {Aziz},
  \citenamefont {Dornheim}, \citenamefont {Azadi}, \citenamefont {Hollebon},
  \citenamefont {Skelt}, \citenamefont {Vinko},\ and\ \citenamefont
  {Gregori}}]{Svensson2024}%
  \BibitemOpen
  \bibfield  {author} {\bibinfo {author} {\bibfnamefont {P.}~\bibnamefont
  {Svensson}}, \bibinfo {author} {\bibfnamefont {Y.}~\bibnamefont {Aziz}},
  \bibinfo {author} {\bibfnamefont {T.}~\bibnamefont {Dornheim}}, \bibinfo
  {author} {\bibfnamefont {S.}~\bibnamefont {Azadi}}, \bibinfo {author}
  {\bibfnamefont {P.}~\bibnamefont {Hollebon}}, \bibinfo {author}
  {\bibfnamefont {A.}~\bibnamefont {Skelt}}, \bibinfo {author} {\bibfnamefont
  {S.~M.}\ \bibnamefont {Vinko}},\ and\ \bibinfo {author} {\bibfnamefont
  {G.}~\bibnamefont {Gregori}},\ }\href {https://arxiv.org/abs/2407.08875}
  {\bibinfo {title} {Modelling of warm dense hydrogen via explicit real time
  electron dynamics: Dynamic structure factors}} (\bibinfo {year}
  {2024}{\natexlab{a}}),\ \Eprint {https://arxiv.org/abs/2407.08875}
  {arXiv:2407.08875 [physics.plasm-ph]} \BibitemShut {NoStop}%
\bibitem [{\citenamefont {Sch\"orner}\ \emph {et~al.}(2022)\citenamefont
  {Sch\"orner}, \citenamefont {R\"uter}, \citenamefont {French},\ and\
  \citenamefont {Redmer}}]{Schorner2022}%
  \BibitemOpen
  \bibfield  {author} {\bibinfo {author} {\bibfnamefont {M.}~\bibnamefont
  {Sch\"orner}}, \bibinfo {author} {\bibfnamefont {H.~R.}\ \bibnamefont
  {R\"uter}}, \bibinfo {author} {\bibfnamefont {M.}~\bibnamefont {French}},\
  and\ \bibinfo {author} {\bibfnamefont {R.}~\bibnamefont {Redmer}},\
  }\bibfield  {title} {\bibinfo {title} {Extending ab initio simulations for
  the ion-ion structure factor of warm dense aluminum to the hydrodynamic limit
  using neural network potentials},\ }\href
  {https://doi.org/10.1103/PhysRevB.105.174310} {\bibfield  {journal} {\bibinfo
   {journal} {Phys. Rev. B}\ }\textbf {\bibinfo {volume} {105}},\ \bibinfo
  {pages} {174310} (\bibinfo {year} {2022})}\BibitemShut {NoStop}%
\bibitem [{\citenamefont {Svensson}\ \emph
  {et~al.}(2024{\natexlab{b}})\citenamefont {Svensson}, \citenamefont
  {Hollebon}, \citenamefont {Plummer}, \citenamefont {Vinko},\ and\
  \citenamefont {Gregori}}]{Svensson2024b}%
  \BibitemOpen
  \bibfield  {author} {\bibinfo {author} {\bibfnamefont {P.}~\bibnamefont
  {Svensson}}, \bibinfo {author} {\bibfnamefont {P.}~\bibnamefont {Hollebon}},
  \bibinfo {author} {\bibfnamefont {D.}~\bibnamefont {Plummer}}, \bibinfo
  {author} {\bibfnamefont {S.~M.}\ \bibnamefont {Vinko}},\ and\ \bibinfo
  {author} {\bibfnamefont {G.}~\bibnamefont {Gregori}},\ }\href
  {https://arxiv.org/abs/2410.08664} {\bibinfo {title} {Modelling of warm dense
  hydrogen via explicit real time electron dynamics: Electron transport
  properties}} (\bibinfo {year} {2024}{\natexlab{b}}),\ \Eprint
  {https://arxiv.org/abs/2410.08664} {arXiv:2410.08664 [physics.plasm-ph]}
  \BibitemShut {NoStop}%
\bibitem [{\citenamefont {Gigosos}\ \emph {et~al.}(2018)\citenamefont
  {Gigosos}, \citenamefont {Gonz\'alez-Herrero}, \citenamefont {Lara},
  \citenamefont {Florido}, \citenamefont {Calisti}, \citenamefont {Ferri},\
  and\ \citenamefont {Talin}}]{Gigosos2018}%
  \BibitemOpen
  \bibfield  {author} {\bibinfo {author} {\bibfnamefont {M.~A.}\ \bibnamefont
  {Gigosos}}, \bibinfo {author} {\bibfnamefont {D.}~\bibnamefont
  {Gonz\'alez-Herrero}}, \bibinfo {author} {\bibfnamefont {N.}~\bibnamefont
  {Lara}}, \bibinfo {author} {\bibfnamefont {R.}~\bibnamefont {Florido}},
  \bibinfo {author} {\bibfnamefont {A.}~\bibnamefont {Calisti}}, \bibinfo
  {author} {\bibfnamefont {S.}~\bibnamefont {Ferri}},\ and\ \bibinfo {author}
  {\bibfnamefont {B.}~\bibnamefont {Talin}},\ }\bibfield  {title} {\bibinfo
  {title} {Classical molecular dynamics simulations of hydrogen plasmas and
  development of an analytical statistical model for computational validity
  assessment},\ }\href {https://doi.org/10.1103/PhysRevE.98.033307} {\bibfield
  {journal} {\bibinfo  {journal} {Phys. Rev. E}\ }\textbf {\bibinfo {volume}
  {98}},\ \bibinfo {pages} {033307} (\bibinfo {year} {2018})}\BibitemShut
  {NoStop}%
\bibitem [{\citenamefont {Saumon}\ and\ \citenamefont
  {Chabrier}(1992)}]{Saumon1992}%
  \BibitemOpen
  \bibfield  {author} {\bibinfo {author} {\bibfnamefont {D.}~\bibnamefont
  {Saumon}}\ and\ \bibinfo {author} {\bibfnamefont {G.}~\bibnamefont
  {Chabrier}},\ }\bibfield  {title} {\bibinfo {title} {Fluid hydrogen at high
  density: Pressure ionization},\ }\href
  {https://doi.org/10.1103/PhysRevA.46.2084} {\bibfield  {journal} {\bibinfo
  {journal} {Physical Review A}\ }\textbf {\bibinfo {volume} {46}},\ \bibinfo
  {pages} {2084} (\bibinfo {year} {1992})}\BibitemShut {NoStop}%
\bibitem [{\citenamefont {Juranek}\ and\ \citenamefont
  {Redmer}(2000)}]{Juranek2000}%
  \BibitemOpen
  \bibfield  {author} {\bibinfo {author} {\bibfnamefont {H.}~\bibnamefont
  {Juranek}}\ and\ \bibinfo {author} {\bibfnamefont {R.}~\bibnamefont
  {Redmer}},\ }\bibfield  {title} {\bibinfo {title} {{Self-consistent fluid
  variational theory for pressure dissociation in dense hydrogen}},\ }\href
  {https://doi.org/10.1063/1.480939} {\bibfield  {journal} {\bibinfo  {journal}
  {The Journal of Chemical Physics}\ }\textbf {\bibinfo {volume} {112}},\
  \bibinfo {pages} {3780} (\bibinfo {year} {2000})}\BibitemShut {NoStop}%
\bibitem [{\citenamefont {Ebeling}\ \emph {et~al.}(2017)\citenamefont
  {Ebeling}, \citenamefont {Fortov},\ and\ \citenamefont
  {Filinov}}]{Ebeling2017}%
  \BibitemOpen
  \bibfield  {author} {\bibinfo {author} {\bibfnamefont {W.}~\bibnamefont
  {Ebeling}}, \bibinfo {author} {\bibfnamefont {V.~E.}\ \bibnamefont
  {Fortov}},\ and\ \bibinfo {author} {\bibfnamefont {V.}~\bibnamefont
  {Filinov}},\ }\href {https://doi.org/10.1007/978-3-319-66637-2} {\emph
  {\bibinfo {title} {Quantum Statistics of Dense Gases and Nonideal Plasmas}}}\
  (\bibinfo  {publisher} {Springer International Publishing},\ \bibinfo {year}
  {2017})\BibitemShut {NoStop}%
\bibitem [{\citenamefont {Winisdoerffer}\ and\ \citenamefont
  {Chabrier}(2005)}]{Winisdoerffer2005}%
  \BibitemOpen
  \bibfield  {author} {\bibinfo {author} {\bibfnamefont {C.}~\bibnamefont
  {Winisdoerffer}}\ and\ \bibinfo {author} {\bibfnamefont {G.}~\bibnamefont
  {Chabrier}},\ }\bibfield  {title} {\bibinfo {title} {Free-energy model for
  fluid helium at high density},\ }\href
  {https://doi.org/10.1103/PhysRevE.71.026402} {\bibfield  {journal} {\bibinfo
  {journal} {Phys. Rev. E}\ }\textbf {\bibinfo {volume} {71}},\ \bibinfo
  {pages} {026402} (\bibinfo {year} {2005})}\BibitemShut {NoStop}%
\bibitem [{\citenamefont {Davletov}\ \emph {et~al.}(2023)\citenamefont
  {Davletov}, \citenamefont {Arkhipov}, \citenamefont {Mukhametkarimov},
  \citenamefont {Yerimbetova},\ and\ \citenamefont {Tkachenko}}]{Davletov2023}%
  \BibitemOpen
  \bibfield  {author} {\bibinfo {author} {\bibfnamefont {A.~E.}\ \bibnamefont
  {Davletov}}, \bibinfo {author} {\bibfnamefont {Y.~V.}\ \bibnamefont
  {Arkhipov}}, \bibinfo {author} {\bibfnamefont {Y.~S.}\ \bibnamefont
  {Mukhametkarimov}}, \bibinfo {author} {\bibfnamefont {L.~T.}\ \bibnamefont
  {Yerimbetova}},\ and\ \bibinfo {author} {\bibfnamefont {I.~M.}\ \bibnamefont
  {Tkachenko}},\ }\bibfield  {title} {\bibinfo {title} {Generalized chemical
  model for plasmas with application to the ionization potential depression},\
  }\href {https://doi.org/10.1088/1367-2630/acd445} {\bibfield  {journal}
  {\bibinfo  {journal} {New Journal of Physics}\ }\textbf {\bibinfo {volume}
  {25}},\ \bibinfo {pages} {063019} (\bibinfo {year} {2023})}\BibitemShut
  {NoStop}%
\bibitem [{\citenamefont {Zimmerman}\ and\ \citenamefont
  {More}(1980)}]{Zimmerman1980}%
  \BibitemOpen
  \bibfield  {author} {\bibinfo {author} {\bibfnamefont {G.}~\bibnamefont
  {Zimmerman}}\ and\ \bibinfo {author} {\bibfnamefont {R.}~\bibnamefont
  {More}},\ }\bibfield  {title} {\bibinfo {title} {Pressure ionization in
  laser-fusion target simulation},\ }\href
  {https://doi.org/https://doi.org/10.1016/0022-4073(80)90055-2} {\bibfield
  {journal} {\bibinfo  {journal} {Journal of Quantitative Spectroscopy and
  Radiative Transfer}\ }\textbf {\bibinfo {volume} {23}},\ \bibinfo {pages}
  {517} (\bibinfo {year} {1980})}\BibitemShut {NoStop}%
\bibitem [{\citenamefont {{Ebeling, W.}}\ \emph {et~al.}(2003)\citenamefont
  {{Ebeling, W.}}, \citenamefont {{Hache, H.}},\ and\ \citenamefont {{Spahn,
  M.}}}]{Ebeling2003}%
  \BibitemOpen
  \bibfield  {author} {\bibinfo {author} {\bibnamefont {{Ebeling, W.}}},
  \bibinfo {author} {\bibnamefont {{Hache, H.}}},\ and\ \bibinfo {author}
  {\bibnamefont {{Spahn, M.}}},\ }\bibfield  {title} {\bibinfo {title}
  {Thermodynamics of ionization and dissociation in hydrogen plasmas including
  fluctuations and magnetic fields},\ }\href
  {https://doi.org/10.1140/epjd/e2003-00041-9} {\bibfield  {journal} {\bibinfo
  {journal} {Eur. Phys. J. D}\ }\textbf {\bibinfo {volume} {23}},\ \bibinfo
  {pages} {265} (\bibinfo {year} {2003})}\BibitemShut {NoStop}%
\bibitem [{\citenamefont {Hummer}\ and\ \citenamefont
  {Mihalas}(1988)}]{Hummer1988}%
  \BibitemOpen
  \bibfield  {author} {\bibinfo {author} {\bibfnamefont {D.}~\bibnamefont
  {Hummer}}\ and\ \bibinfo {author} {\bibfnamefont {D.}~\bibnamefont
  {Mihalas}},\ }\bibfield  {title} {\bibinfo {title} {The equation of state for
  stellar envelopes. {I} - an occupation probability formalism for the
  truncation of internal partition functions},\ }\href
  {https://doi.org/10.1086/166600} {\bibfield  {journal} {\bibinfo  {journal}
  {The Astrophysical Journal}\ }\textbf {\bibinfo {volume} {331}},\ \bibinfo
  {pages} {794} (\bibinfo {year} {1988})}\BibitemShut {NoStop}%
\bibitem [{\citenamefont {Potekhin}(1996)}]{Potekhin1996}%
  \BibitemOpen
  \bibfield  {author} {\bibinfo {author} {\bibfnamefont {A.~Y.}\ \bibnamefont
  {Potekhin}},\ }\bibfield  {title} {\bibinfo {title} {{Ionization equilibrium
  of hot hydrogen plasma}},\ }\href {https://doi.org/10.1063/1.871547}
  {\bibfield  {journal} {\bibinfo  {journal} {Physics of Plasmas}\ }\textbf
  {\bibinfo {volume} {3}},\ \bibinfo {pages} {4156} (\bibinfo {year}
  {1996})}\BibitemShut {NoStop}%
\bibitem [{\citenamefont {Chihara}(1987)}]{Chihara1987}%
  \BibitemOpen
  \bibfield  {author} {\bibinfo {author} {\bibfnamefont {J.}~\bibnamefont
  {Chihara}},\ }\bibfield  {title} {\bibinfo {title} {Difference in x-ray
  scattering between metallic and non-metallic liquids due to conduction
  electrons},\ }\href {https://doi.org/10.1088/0305-4608/17/2/002} {\bibfield
  {journal} {\bibinfo  {journal} {Journal of Physics F: Metal Physics}\
  }\textbf {\bibinfo {volume} {17}},\ \bibinfo {pages} {295} (\bibinfo {year}
  {1987})}\BibitemShut {NoStop}%
\bibitem [{\citenamefont {Stanton}\ and\ \citenamefont
  {Murillo}(2016)}]{Stanton2016}%
  \BibitemOpen
  \bibfield  {author} {\bibinfo {author} {\bibfnamefont {L.~G.}\ \bibnamefont
  {Stanton}}\ and\ \bibinfo {author} {\bibfnamefont {M.~S.}\ \bibnamefont
  {Murillo}},\ }\bibfield  {title} {\bibinfo {title} {Ionic transport in
  high-energy-density matter},\ }\href
  {https://doi.org/10.1103/PhysRevE.93.043203} {\bibfield  {journal} {\bibinfo
  {journal} {Phys. Rev. E}\ }\textbf {\bibinfo {volume} {93}},\ \bibinfo
  {pages} {043203} (\bibinfo {year} {2016})}\BibitemShut {NoStop}%
\bibitem [{\citenamefont {Chung}\ \emph {et~al.}(2005)\citenamefont {Chung},
  \citenamefont {Chen}, \citenamefont {Morgan}, \citenamefont {Ralchenko},\
  and\ \citenamefont {Lee}}]{Chung2005}%
  \BibitemOpen
  \bibfield  {author} {\bibinfo {author} {\bibfnamefont {H.-K.}\ \bibnamefont
  {Chung}}, \bibinfo {author} {\bibfnamefont {M.}~\bibnamefont {Chen}},
  \bibinfo {author} {\bibfnamefont {W.}~\bibnamefont {Morgan}}, \bibinfo
  {author} {\bibfnamefont {Y.}~\bibnamefont {Ralchenko}},\ and\ \bibinfo
  {author} {\bibfnamefont {R.}~\bibnamefont {Lee}},\ }\bibfield  {title}
  {\bibinfo {title} {Flychk: Generalized population kinetics and spectral model
  for rapid spectroscopic analysis for all elements},\ }\href
  {https://doi.org/https://doi.org/10.1016/j.hedp.2005.07.001} {\bibfield
  {journal} {\bibinfo  {journal} {High Energy Density Physics}\ }\textbf
  {\bibinfo {volume} {1}},\ \bibinfo {pages} {3} (\bibinfo {year}
  {2005})}\BibitemShut {NoStop}%
\bibitem [{\citenamefont {Callow}\ \emph {et~al.}(2023)\citenamefont {Callow},
  \citenamefont {Kraisler},\ and\ \citenamefont {Cangi}}]{Callow2023}%
  \BibitemOpen
  \bibfield  {author} {\bibinfo {author} {\bibfnamefont {T.~J.}\ \bibnamefont
  {Callow}}, \bibinfo {author} {\bibfnamefont {E.}~\bibnamefont {Kraisler}},\
  and\ \bibinfo {author} {\bibfnamefont {A.}~\bibnamefont {Cangi}},\ }\bibfield
   {title} {\bibinfo {title} {Improved calculations of mean ionization states
  with an average-atom model},\ }\href
  {https://doi.org/10.1103/PhysRevResearch.5.013049} {\bibfield  {journal}
  {\bibinfo  {journal} {Phys. Rev. Res.}\ }\textbf {\bibinfo {volume} {5}},\
  \bibinfo {pages} {013049} (\bibinfo {year} {2023})}\BibitemShut {NoStop}%
\bibitem [{\citenamefont {Ciricosta}\ \emph {et~al.}(2012)\citenamefont
  {Ciricosta}, \citenamefont {Vinko}, \citenamefont {Chung}, \citenamefont
  {Cho}, \citenamefont {Brown}, \citenamefont {Burian}, \citenamefont
  {Chalupsk\'y}, \citenamefont {Engelhorn}, \citenamefont {Falcone},
  \citenamefont {Graves}, \citenamefont {H\'ajkov\'a}, \citenamefont
  {Higginbotham}, \citenamefont {Juha}, \citenamefont {Krzywinski},
  \citenamefont {Lee}, \citenamefont {Messerschmidt}, \citenamefont {Murphy},
  \citenamefont {Ping}, \citenamefont {Rackstraw}, \citenamefont {Scherz},
  \citenamefont {Schlotter}, \citenamefont {Toleikis}, \citenamefont {Turner},
  \citenamefont {Vysin}, \citenamefont {Wang}, \citenamefont {Wu},
  \citenamefont {Zastrau}, \citenamefont {Zhu}, \citenamefont {Lee},
  \citenamefont {Heimann}, \citenamefont {Nagler},\ and\ \citenamefont
  {Wark}}]{Ciricosta2012}%
  \BibitemOpen
  \bibfield  {author} {\bibinfo {author} {\bibfnamefont {O.}~\bibnamefont
  {Ciricosta}}, \bibinfo {author} {\bibfnamefont {S.~M.}\ \bibnamefont
  {Vinko}}, \bibinfo {author} {\bibfnamefont {H.-K.}\ \bibnamefont {Chung}},
  \bibinfo {author} {\bibfnamefont {B.-I.}\ \bibnamefont {Cho}}, \bibinfo
  {author} {\bibfnamefont {C.~R.~D.}\ \bibnamefont {Brown}}, \bibinfo {author}
  {\bibfnamefont {T.}~\bibnamefont {Burian}}, \bibinfo {author} {\bibfnamefont
  {J.}~\bibnamefont {Chalupsk\'y}}, \bibinfo {author} {\bibfnamefont
  {K.}~\bibnamefont {Engelhorn}}, \bibinfo {author} {\bibfnamefont {R.~W.}\
  \bibnamefont {Falcone}}, \bibinfo {author} {\bibfnamefont {C.}~\bibnamefont
  {Graves}}, \bibinfo {author} {\bibfnamefont {V.}~\bibnamefont {H\'ajkov\'a}},
  \bibinfo {author} {\bibfnamefont {A.}~\bibnamefont {Higginbotham}}, \bibinfo
  {author} {\bibfnamefont {L.}~\bibnamefont {Juha}}, \bibinfo {author}
  {\bibfnamefont {J.}~\bibnamefont {Krzywinski}}, \bibinfo {author}
  {\bibfnamefont {H.~J.}\ \bibnamefont {Lee}}, \bibinfo {author} {\bibfnamefont
  {M.}~\bibnamefont {Messerschmidt}}, \bibinfo {author} {\bibfnamefont {C.~D.}\
  \bibnamefont {Murphy}}, \bibinfo {author} {\bibfnamefont {Y.}~\bibnamefont
  {Ping}}, \bibinfo {author} {\bibfnamefont {D.~S.}\ \bibnamefont {Rackstraw}},
  \bibinfo {author} {\bibfnamefont {A.}~\bibnamefont {Scherz}}, \bibinfo
  {author} {\bibfnamefont {W.}~\bibnamefont {Schlotter}}, \bibinfo {author}
  {\bibfnamefont {S.}~\bibnamefont {Toleikis}}, \bibinfo {author}
  {\bibfnamefont {J.~J.}\ \bibnamefont {Turner}}, \bibinfo {author}
  {\bibfnamefont {L.}~\bibnamefont {Vysin}}, \bibinfo {author} {\bibfnamefont
  {T.}~\bibnamefont {Wang}}, \bibinfo {author} {\bibfnamefont {B.}~\bibnamefont
  {Wu}}, \bibinfo {author} {\bibfnamefont {U.}~\bibnamefont {Zastrau}},
  \bibinfo {author} {\bibfnamefont {D.}~\bibnamefont {Zhu}}, \bibinfo {author}
  {\bibfnamefont {R.~W.}\ \bibnamefont {Lee}}, \bibinfo {author} {\bibfnamefont
  {P.}~\bibnamefont {Heimann}}, \bibinfo {author} {\bibfnamefont
  {B.}~\bibnamefont {Nagler}},\ and\ \bibinfo {author} {\bibfnamefont {J.~S.}\
  \bibnamefont {Wark}},\ }\bibfield  {title} {\bibinfo {title} {Direct
  measurements of the ionization potential depression in a dense plasma},\
  }\href {https://doi.org/10.1103/PhysRevLett.109.065002} {\bibfield  {journal}
  {\bibinfo  {journal} {Phys. Rev. Lett.}\ }\textbf {\bibinfo {volume} {109}},\
  \bibinfo {pages} {065002} (\bibinfo {year} {2012})}\BibitemShut {NoStop}%
\bibitem [{\citenamefont {Frenkel}\ and\ \citenamefont
  {Smit}(2002)}]{Frenkel2002}%
  \BibitemOpen
  \bibfield  {author} {\bibinfo {author} {\bibfnamefont {D.}~\bibnamefont
  {Frenkel}}\ and\ \bibinfo {author} {\bibfnamefont {B.}~\bibnamefont {Smit}},\
  }\bibfield  {title} {\bibinfo {title} {Chapter 7 - free energy
  calculations},\ }in\ \href
  {https://doi.org/https://doi.org/10.1016/B978-012267351-1/50009-2} {\emph
  {\bibinfo {booktitle} {Understanding Molecular Simulation (Second
  Edition)}}},\ \bibinfo {editor} {edited by\ \bibinfo {editor} {\bibfnamefont
  {D.}~\bibnamefont {Frenkel}}\ and\ \bibinfo {editor} {\bibfnamefont
  {B.}~\bibnamefont {Smit}}}\ (\bibinfo  {publisher} {Academic Press},\
  \bibinfo {address} {San Diego},\ \bibinfo {year} {2002})\ \bibinfo {edition}
  {second edition}\ ed.,\ pp.\ \bibinfo {pages} {167--200}\BibitemShut
  {NoStop}%
\bibitem [{\citenamefont {Fraser}\ \emph {et~al.}(1996)\citenamefont {Fraser},
  \citenamefont {Foulkes}, \citenamefont {Rajagopal}, \citenamefont {Needs},
  \citenamefont {Kenny},\ and\ \citenamefont {Williamson}}]{Fraser1996}%
  \BibitemOpen
  \bibfield  {author} {\bibinfo {author} {\bibfnamefont {L.~M.}\ \bibnamefont
  {Fraser}}, \bibinfo {author} {\bibfnamefont {W.~M.~C.}\ \bibnamefont
  {Foulkes}}, \bibinfo {author} {\bibfnamefont {G.}~\bibnamefont {Rajagopal}},
  \bibinfo {author} {\bibfnamefont {R.~J.}\ \bibnamefont {Needs}}, \bibinfo
  {author} {\bibfnamefont {S.~D.}\ \bibnamefont {Kenny}},\ and\ \bibinfo
  {author} {\bibfnamefont {A.~J.}\ \bibnamefont {Williamson}},\ }\bibfield
  {title} {\bibinfo {title} {Finite-size effects and coulomb interactions in
  quantum monte carlo calculations for homogeneous systems with periodic
  boundary conditions},\ }\href {https://doi.org/10.1103/PhysRevB.53.1814}
  {\bibfield  {journal} {\bibinfo  {journal} {Phys. Rev. B}\ }\textbf {\bibinfo
  {volume} {53}},\ \bibinfo {pages} {1814} (\bibinfo {year}
  {1996})}\BibitemShut {NoStop}%
\bibitem [{\citenamefont {Rapaport}(2004)}]{Rapaport2004}%
  \BibitemOpen
  \bibfield  {author} {\bibinfo {author} {\bibfnamefont {D.~C.}\ \bibnamefont
  {Rapaport}},\ }\href@noop {} {\emph {\bibinfo {title} {The Art of Molecular
  Dynamics Simulation}}},\ \bibinfo {edition} {2nd}\ ed.\ (\bibinfo
  {publisher} {Cambridge University Press},\ \bibinfo {year}
  {2004})\BibitemShut {NoStop}%
\bibitem [{\citenamefont {Petersen}(1995)}]{Peterson1995}%
  \BibitemOpen
  \bibfield  {author} {\bibinfo {author} {\bibfnamefont {H.~G.}\ \bibnamefont
  {Petersen}},\ }\bibfield  {title} {\bibinfo {title} {{Accuracy and efficiency
  of the particle mesh Ewald method}},\ }\href
  {https://doi.org/10.1063/1.470043} {\bibfield  {journal} {\bibinfo  {journal}
  {The Journal of Chemical Physics}\ }\textbf {\bibinfo {volume} {103}},\
  \bibinfo {pages} {3668} (\bibinfo {year} {1995})}\BibitemShut {NoStop}%
\bibitem [{\citenamefont {W\"unsch}\ \emph {et~al.}(2009)\citenamefont
  {W\"unsch}, \citenamefont {Vorberger},\ and\ \citenamefont
  {Gericke}}]{Wunsch2009}%
  \BibitemOpen
  \bibfield  {author} {\bibinfo {author} {\bibfnamefont {K.}~\bibnamefont
  {W\"unsch}}, \bibinfo {author} {\bibfnamefont {J.}~\bibnamefont
  {Vorberger}},\ and\ \bibinfo {author} {\bibfnamefont {D.~O.}\ \bibnamefont
  {Gericke}},\ }\bibfield  {title} {\bibinfo {title} {Ion structure in warm
  dense matter: Benchmarking solutions of hypernetted-chain equations by
  first-principle simulations},\ }\href
  {https://doi.org/10.1103/PhysRevE.79.010201} {\bibfield  {journal} {\bibinfo
  {journal} {Phys. Rev. E}\ }\textbf {\bibinfo {volume} {79}},\ \bibinfo
  {pages} {010201(R)} (\bibinfo {year} {2009})}\BibitemShut {NoStop}%
\bibitem [{\citenamefont {Vorberger}\ and\ \citenamefont
  {Gericke}(2013)}]{Vorberger2013}%
  \BibitemOpen
  \bibfield  {author} {\bibinfo {author} {\bibfnamefont {J.}~\bibnamefont
  {Vorberger}}\ and\ \bibinfo {author} {\bibfnamefont {D.}~\bibnamefont
  {Gericke}},\ }\bibfield  {title} {\bibinfo {title} {Effective ion–ion
  potentials in warm dense matter},\ }\href
  {https://doi.org/https://doi.org/10.1016/j.hedp.2012.12.009} {\bibfield
  {journal} {\bibinfo  {journal} {High Energy Density Physics}\ }\textbf
  {\bibinfo {volume} {9}},\ \bibinfo {pages} {178} (\bibinfo {year}
  {2013})}\BibitemShut {NoStop}%
\bibitem [{\citenamefont {Thompson}\ \emph {et~al.}(2022)\citenamefont
  {Thompson}, \citenamefont {Aktulga}, \citenamefont {Berger}, \citenamefont
  {Bolintineanu}, \citenamefont {Brown}, \citenamefont {Crozier}, \citenamefont
  {in~'t Veld}, \citenamefont {Kohlmeyer}, \citenamefont {Moore}, \citenamefont
  {Nguyen}, \citenamefont {Shan}, \citenamefont {Stevens}, \citenamefont
  {Tranchida}, \citenamefont {Trott},\ and\ \citenamefont {Plimpton}}]{LAMMPS}%
  \BibitemOpen
  \bibfield  {author} {\bibinfo {author} {\bibfnamefont {A.~P.}\ \bibnamefont
  {Thompson}}, \bibinfo {author} {\bibfnamefont {H.~M.}\ \bibnamefont
  {Aktulga}}, \bibinfo {author} {\bibfnamefont {R.}~\bibnamefont {Berger}},
  \bibinfo {author} {\bibfnamefont {D.~S.}\ \bibnamefont {Bolintineanu}},
  \bibinfo {author} {\bibfnamefont {W.~M.}\ \bibnamefont {Brown}}, \bibinfo
  {author} {\bibfnamefont {P.~S.}\ \bibnamefont {Crozier}}, \bibinfo {author}
  {\bibfnamefont {P.~J.}\ \bibnamefont {in~'t Veld}}, \bibinfo {author}
  {\bibfnamefont {A.}~\bibnamefont {Kohlmeyer}}, \bibinfo {author}
  {\bibfnamefont {S.~G.}\ \bibnamefont {Moore}}, \bibinfo {author}
  {\bibfnamefont {T.~D.}\ \bibnamefont {Nguyen}}, \bibinfo {author}
  {\bibfnamefont {R.}~\bibnamefont {Shan}}, \bibinfo {author} {\bibfnamefont
  {M.~J.}\ \bibnamefont {Stevens}}, \bibinfo {author} {\bibfnamefont
  {J.}~\bibnamefont {Tranchida}}, \bibinfo {author} {\bibfnamefont
  {C.}~\bibnamefont {Trott}},\ and\ \bibinfo {author} {\bibfnamefont {S.~J.}\
  \bibnamefont {Plimpton}},\ }\bibfield  {title} {\bibinfo {title} {{LAMMPS} -
  a flexible simulation tool for particle-based materials modeling at the
  atomic, meso, and continuum scales},\ }\href
  {https://doi.org/10.1016/j.cpc.2021.108171} {\bibfield  {journal} {\bibinfo
  {journal} {Comp. Phys. Comm.}\ }\textbf {\bibinfo {volume} {271}},\ \bibinfo
  {pages} {108171} (\bibinfo {year} {2022})}\BibitemShut {NoStop}%
\bibitem [{\citenamefont {Kumar}\ \emph {et~al.}(2021)\citenamefont {Kumar},
  \citenamefont {Poser}, \citenamefont {Sch\"ottler}, \citenamefont
  {Kleinschmidt}, \citenamefont {Dietrich}, \citenamefont {Wicht},
  \citenamefont {French},\ and\ \citenamefont {Redmer}}]{Kumar2021}%
  \BibitemOpen
  \bibfield  {author} {\bibinfo {author} {\bibfnamefont {S.}~\bibnamefont
  {Kumar}}, \bibinfo {author} {\bibfnamefont {A.~J.}\ \bibnamefont {Poser}},
  \bibinfo {author} {\bibfnamefont {M.}~\bibnamefont {Sch\"ottler}}, \bibinfo
  {author} {\bibfnamefont {U.}~\bibnamefont {Kleinschmidt}}, \bibinfo {author}
  {\bibfnamefont {W.}~\bibnamefont {Dietrich}}, \bibinfo {author}
  {\bibfnamefont {J.}~\bibnamefont {Wicht}}, \bibinfo {author} {\bibfnamefont
  {M.}~\bibnamefont {French}},\ and\ \bibinfo {author} {\bibfnamefont
  {R.}~\bibnamefont {Redmer}},\ }\bibfield  {title} {\bibinfo {title}
  {Ionization and transport in partially ionized multicomponent plasmas:
  Application to atmospheres of hot jupiters},\ }\href
  {https://doi.org/10.1103/PhysRevE.103.063203} {\bibfield  {journal} {\bibinfo
   {journal} {Phys. Rev. E}\ }\textbf {\bibinfo {volume} {103}},\ \bibinfo
  {pages} {063203} (\bibinfo {year} {2021})}\BibitemShut {NoStop}%
\bibitem [{\citenamefont {Beutler}\ \emph {et~al.}(1994)\citenamefont
  {Beutler}, \citenamefont {Mark}, \citenamefont {{van Schaik}}, \citenamefont
  {Gerber},\ and\ \citenamefont {{van Gunsteren}}}]{Beutler1994}%
  \BibitemOpen
  \bibfield  {author} {\bibinfo {author} {\bibfnamefont {T.~C.}\ \bibnamefont
  {Beutler}}, \bibinfo {author} {\bibfnamefont {A.~E.}\ \bibnamefont {Mark}},
  \bibinfo {author} {\bibfnamefont {R.~C.}\ \bibnamefont {{van Schaik}}},
  \bibinfo {author} {\bibfnamefont {P.~R.}\ \bibnamefont {Gerber}},\ and\
  \bibinfo {author} {\bibfnamefont {W.~F.}\ \bibnamefont {{van Gunsteren}}},\
  }\bibfield  {title} {\bibinfo {title} {Avoiding singularities and numerical
  instabilities in free energy calculations based on molecular simulations},\
  }\href {https://doi.org/https://doi.org/10.1016/0009-2614(94)00397-1}
  {\bibfield  {journal} {\bibinfo  {journal} {Chemical Physics Letters}\
  }\textbf {\bibinfo {volume} {222}},\ \bibinfo {pages} {529} (\bibinfo {year}
  {1994})}\BibitemShut {NoStop}%
\bibitem [{\citenamefont {de~Ruiter}\ \emph {et~al.}(2021)\citenamefont
  {de~Ruiter}, \citenamefont {Petrov},\ and\ \citenamefont
  {Oostenbrink}}]{Ruiter2021}%
  \BibitemOpen
  \bibfield  {author} {\bibinfo {author} {\bibfnamefont {A.}~\bibnamefont
  {de~Ruiter}}, \bibinfo {author} {\bibfnamefont {D.}~\bibnamefont {Petrov}},\
  and\ \bibinfo {author} {\bibfnamefont {C.}~\bibnamefont {Oostenbrink}},\
  }\bibfield  {title} {\bibinfo {title} {Optimization of alchemical pathways
  using extended thermodynamic integration},\ }\href
  {https://doi.org/10.1021/acs.jctc.0c01170} {\bibfield  {journal} {\bibinfo
  {journal} {Journal of Chemical Theory and Computation}\ }\textbf {\bibinfo
  {volume} {17}},\ \bibinfo {pages} {56} (\bibinfo {year} {2021})}\BibitemShut
  {NoStop}%
\bibitem [{\citenamefont {Figueirido}\ \emph {et~al.}(1995)\citenamefont
  {Figueirido}, \citenamefont {Buono},\ and\ \citenamefont
  {Levy}}]{Figueirido1995}%
  \BibitemOpen
  \bibfield  {author} {\bibinfo {author} {\bibfnamefont {F.}~\bibnamefont
  {Figueirido}}, \bibinfo {author} {\bibfnamefont {G.~S.~D.}\ \bibnamefont
  {Buono}},\ and\ \bibinfo {author} {\bibfnamefont {R.~M.}\ \bibnamefont
  {Levy}},\ }\bibfield  {title} {\bibinfo {title} {On finite‐size effects in
  computer simulations using the ewald potential},\ }\href
  {https://doi.org/10.1063/1.470721} {\bibfield  {journal} {\bibinfo  {journal}
  {The Journal of Chemical Physics}\ }\textbf {\bibinfo {volume} {103}},\
  \bibinfo {pages} {6133} (\bibinfo {year} {1995})}\BibitemShut {NoStop}%
\bibitem [{\citenamefont {Caillol}(1999{\natexlab{b}})}]{Caillol1999a}%
  \BibitemOpen
  \bibfield  {author} {\bibinfo {author} {\bibfnamefont {J.~M.}\ \bibnamefont
  {Caillol}},\ }\bibfield  {title} {\bibinfo {title} {{Numerical simulations of
  Coulomb systems: A comparison between hyperspherical and periodic boundary
  conditions}},\ }\href {https://doi.org/10.1063/1.479947} {\bibfield
  {journal} {\bibinfo  {journal} {The Journal of Chemical Physics}\ }\textbf
  {\bibinfo {volume} {111}},\ \bibinfo {pages} {6528} (\bibinfo {year}
  {1999}{\natexlab{b}})}\BibitemShut {NoStop}%
\bibitem [{\citenamefont {Caillol}\ and\ \citenamefont
  {Gilles}(2000)}]{Caillol2000}%
  \BibitemOpen
  \bibfield  {author} {\bibinfo {author} {\bibfnamefont {J.~M.}\ \bibnamefont
  {Caillol}}\ and\ \bibinfo {author} {\bibfnamefont {D.}~\bibnamefont
  {Gilles}},\ }\bibfield  {title} {\bibinfo {title} {Monte carlo simulations of
  the yukawa one-component plasma},\ }\href
  {https://doi.org/10.1023/A:1018727428374/METRICS} {\bibfield  {journal}
  {\bibinfo  {journal} {Journal of Statistical Physics}\ }\textbf {\bibinfo
  {volume} {100}},\ \bibinfo {pages} {933} (\bibinfo {year}
  {2000})}\BibitemShut {NoStop}%
\bibitem [{\citenamefont {Moldabekov}\ \emph {et~al.}(2022)\citenamefont
  {Moldabekov}, \citenamefont {Dornheim},\ and\ \citenamefont
  {Bonitz}}]{Moldabekov2022}%
  \BibitemOpen
  \bibfield  {author} {\bibinfo {author} {\bibfnamefont {Z.~A.}\ \bibnamefont
  {Moldabekov}}, \bibinfo {author} {\bibfnamefont {T.}~\bibnamefont
  {Dornheim}},\ and\ \bibinfo {author} {\bibfnamefont {M.}~\bibnamefont
  {Bonitz}},\ }\bibfield  {title} {\bibinfo {title} {Screening of a test charge
  in a free-electron gas at warm dense matter and dense non-ideal plasma
  conditions},\ }\href {https://doi.org/https://doi.org/10.1002/ctpp.202000176}
  {\bibfield  {journal} {\bibinfo  {journal} {Contributions to Plasma Physics}\
  }\textbf {\bibinfo {volume} {62}},\ \bibinfo {pages} {e202000176} (\bibinfo
  {year} {2022})}\BibitemShut {NoStop}%
\bibitem [{\citenamefont {Gawne}\ \emph {et~al.}(2023)\citenamefont {Gawne},
  \citenamefont {Campbell}, \citenamefont {Forte}, \citenamefont {Hollebon},
  \citenamefont {Perez-Callejo}, \citenamefont {Humphries}, \citenamefont
  {Karnbach}, \citenamefont {Kasim}, \citenamefont {Preston}, \citenamefont
  {Lee}, \citenamefont {Miscampbell}, \citenamefont {van~den Berg},
  \citenamefont {Nagler}, \citenamefont {Ren}, \citenamefont {Royle},
  \citenamefont {Wark},\ and\ \citenamefont {Vinko}}]{Gawne2023}%
  \BibitemOpen
  \bibfield  {author} {\bibinfo {author} {\bibfnamefont {T.}~\bibnamefont
  {Gawne}}, \bibinfo {author} {\bibfnamefont {T.}~\bibnamefont {Campbell}},
  \bibinfo {author} {\bibfnamefont {A.}~\bibnamefont {Forte}}, \bibinfo
  {author} {\bibfnamefont {P.}~\bibnamefont {Hollebon}}, \bibinfo {author}
  {\bibfnamefont {G.}~\bibnamefont {Perez-Callejo}}, \bibinfo {author}
  {\bibfnamefont {O.~S.}\ \bibnamefont {Humphries}}, \bibinfo {author}
  {\bibfnamefont {O.}~\bibnamefont {Karnbach}}, \bibinfo {author}
  {\bibfnamefont {M.~F.}\ \bibnamefont {Kasim}}, \bibinfo {author}
  {\bibfnamefont {T.~R.}\ \bibnamefont {Preston}}, \bibinfo {author}
  {\bibfnamefont {H.~J.}\ \bibnamefont {Lee}}, \bibinfo {author} {\bibfnamefont
  {A.}~\bibnamefont {Miscampbell}}, \bibinfo {author} {\bibfnamefont {Q.~Y.}\
  \bibnamefont {van~den Berg}}, \bibinfo {author} {\bibfnamefont
  {B.}~\bibnamefont {Nagler}}, \bibinfo {author} {\bibfnamefont
  {S.}~\bibnamefont {Ren}}, \bibinfo {author} {\bibfnamefont {R.~B.}\
  \bibnamefont {Royle}}, \bibinfo {author} {\bibfnamefont {J.~S.}\ \bibnamefont
  {Wark}},\ and\ \bibinfo {author} {\bibfnamefont {S.~M.}\ \bibnamefont
  {Vinko}},\ }\bibfield  {title} {\bibinfo {title} {Investigating mechanisms of
  state localization in highly ionized dense plasmas},\ }\href
  {https://doi.org/10.1103/PhysRevE.108.035210} {\bibfield  {journal} {\bibinfo
   {journal} {Phys. Rev. E}\ }\textbf {\bibinfo {volume} {108}},\ \bibinfo
  {pages} {035210} (\bibinfo {year} {2023})}\BibitemShut {NoStop}%
\bibitem [{\citenamefont {Gawne}\ \emph {et~al.}(2024)\citenamefont {Gawne},
  \citenamefont {Vinko},\ and\ \citenamefont {Wark}}]{Gawne2024}%
  \BibitemOpen
  \bibfield  {author} {\bibinfo {author} {\bibfnamefont {T.}~\bibnamefont
  {Gawne}}, \bibinfo {author} {\bibfnamefont {S.~M.}\ \bibnamefont {Vinko}},\
  and\ \bibinfo {author} {\bibfnamefont {J.~S.}\ \bibnamefont {Wark}},\
  }\bibfield  {title} {\bibinfo {title} {Quantifying ionization in hot dense
  plasmas},\ }\href {https://doi.org/10.1103/PhysRevE.109.L023201} {\bibfield
  {journal} {\bibinfo  {journal} {Physical Review E}\ }\textbf {\bibinfo
  {volume} {109}},\ \bibinfo {pages} {023201} (\bibinfo {year}
  {2024})}\BibitemShut {NoStop}%
\bibitem [{\citenamefont {Bethkenhagen}\ \emph {et~al.}(2020)\citenamefont
  {Bethkenhagen}, \citenamefont {Witte}, \citenamefont {Sch\"orner},
  \citenamefont {R\"opke}, \citenamefont {D\"oppner}, \citenamefont {Kraus},
  \citenamefont {Glenzer}, \citenamefont {Sterne},\ and\ \citenamefont
  {Redmer}}]{Bethkenhagen2020}%
  \BibitemOpen
  \bibfield  {author} {\bibinfo {author} {\bibfnamefont {M.}~\bibnamefont
  {Bethkenhagen}}, \bibinfo {author} {\bibfnamefont {B.~B.~L.}\ \bibnamefont
  {Witte}}, \bibinfo {author} {\bibfnamefont {M.}~\bibnamefont {Sch\"orner}},
  \bibinfo {author} {\bibfnamefont {G.}~\bibnamefont {R\"opke}}, \bibinfo
  {author} {\bibfnamefont {T.}~\bibnamefont {D\"oppner}}, \bibinfo {author}
  {\bibfnamefont {D.}~\bibnamefont {Kraus}}, \bibinfo {author} {\bibfnamefont
  {S.~H.}\ \bibnamefont {Glenzer}}, \bibinfo {author} {\bibfnamefont {P.~A.}\
  \bibnamefont {Sterne}},\ and\ \bibinfo {author} {\bibfnamefont
  {R.}~\bibnamefont {Redmer}},\ }\bibfield  {title} {\bibinfo {title} {Carbon
  ionization at gigabar pressures: An ab initio perspective on astrophysical
  high-density plasmas},\ }\href
  {https://doi.org/10.1103/PhysRevResearch.2.023260} {\bibfield  {journal}
  {\bibinfo  {journal} {Phys. Rev. Res.}\ }\textbf {\bibinfo {volume} {2}},\
  \bibinfo {pages} {023260} (\bibinfo {year} {2020})}\BibitemShut {NoStop}%
\bibitem [{\citenamefont {Cl\'erouin}\ \emph {et~al.}(2022)\citenamefont
  {Cl\'erouin}, \citenamefont {Blanchet}, \citenamefont {Blancard},
  \citenamefont {Faussurier}, \citenamefont {Soubiran},\ and\ \citenamefont
  {Bethkenhagen}}]{Clerouin2022}%
  \BibitemOpen
  \bibfield  {author} {\bibinfo {author} {\bibfnamefont {J.}~\bibnamefont
  {Cl\'erouin}}, \bibinfo {author} {\bibfnamefont {A.}~\bibnamefont
  {Blanchet}}, \bibinfo {author} {\bibfnamefont {C.}~\bibnamefont {Blancard}},
  \bibinfo {author} {\bibfnamefont {G.}~\bibnamefont {Faussurier}}, \bibinfo
  {author} {\bibfnamefont {F.}~\bibnamefont {Soubiran}},\ and\ \bibinfo
  {author} {\bibfnamefont {M.}~\bibnamefont {Bethkenhagen}},\ }\bibfield
  {title} {\bibinfo {title} {Equivalence between pressure- and
  structure-defined ionization in hot dense carbon},\ }\href
  {https://doi.org/10.1103/PhysRevE.106.045204} {\bibfield  {journal} {\bibinfo
   {journal} {Phys. Rev. E}\ }\textbf {\bibinfo {volume} {106}},\ \bibinfo
  {pages} {045204} (\bibinfo {year} {2022})}\BibitemShut {NoStop}%
\bibitem [{\citenamefont {Ecker}\ and\ \citenamefont
  {Kröll}(1963)}]{Ecker1963}%
  \BibitemOpen
  \bibfield  {author} {\bibinfo {author} {\bibfnamefont {G.}~\bibnamefont
  {Ecker}}\ and\ \bibinfo {author} {\bibfnamefont {W.}~\bibnamefont {Kröll}},\
  }\bibfield  {title} {\bibinfo {title} {{Lowering of the Ionization Energy for
  a Plasma in Thermodynamic Equilibrium}},\ }\href
  {https://doi.org/10.1063/1.1724509} {\bibfield  {journal} {\bibinfo
  {journal} {The Physics of Fluids}\ }\textbf {\bibinfo {volume} {6}},\
  \bibinfo {pages} {62} (\bibinfo {year} {1963})}\BibitemShut {NoStop}%
\bibitem [{\citenamefont {Griem}(1962)}]{Griem1962}%
  \BibitemOpen
  \bibfield  {author} {\bibinfo {author} {\bibfnamefont {H.~R.}\ \bibnamefont
  {Griem}},\ }\bibfield  {title} {\bibinfo {title} {High-density corrections in
  plasma spectroscopy},\ }\href {https://doi.org/10.1103/PhysRev.128.997}
  {\bibfield  {journal} {\bibinfo  {journal} {Phys. Rev.}\ }\textbf {\bibinfo
  {volume} {128}},\ \bibinfo {pages} {997} (\bibinfo {year}
  {1962})}\BibitemShut {NoStop}%
\bibitem [{\citenamefont {Potekhin}\ and\ \citenamefont
  {Chabrier}(2000)}]{Potekhin2000}%
  \BibitemOpen
  \bibfield  {author} {\bibinfo {author} {\bibfnamefont {A.~Y.}\ \bibnamefont
  {Potekhin}}\ and\ \bibinfo {author} {\bibfnamefont {G.}~\bibnamefont
  {Chabrier}},\ }\bibfield  {title} {\bibinfo {title} {Equation of state of
  fully ionized electron-ion plasmas. {II}. extension to relativistic densities
  and to the solid phase},\ }\href {https://doi.org/10.1103/PhysRevE.62.8554}
  {\bibfield  {journal} {\bibinfo  {journal} {Phys. Rev. E}\ }\textbf {\bibinfo
  {volume} {62}},\ \bibinfo {pages} {8554} (\bibinfo {year}
  {2000})}\BibitemShut {NoStop}%
\bibitem [{\citenamefont {Carnahan}\ and\ \citenamefont
  {Starling}(1969)}]{Carnahan1969}%
  \BibitemOpen
  \bibfield  {author} {\bibinfo {author} {\bibfnamefont {N.~F.}\ \bibnamefont
  {Carnahan}}\ and\ \bibinfo {author} {\bibfnamefont {K.~E.}\ \bibnamefont
  {Starling}},\ }\bibfield  {title} {\bibinfo {title} {{Equation of State for
  Nonattracting Rigid Spheres}},\ }\href {https://doi.org/10.1063/1.1672048}
  {\bibfield  {journal} {\bibinfo  {journal} {The Journal of Chemical Physics}\
  }\textbf {\bibinfo {volume} {51}},\ \bibinfo {pages} {635} (\bibinfo {year}
  {1969})}\BibitemShut {NoStop}%
\end{thebibliography}%

\end{document}